\documentclass[10pt, conference]{IEEEtran}

\usepackage[T1]{fontenc}

\usepackage[caption=false]{subfig}

\usepackage{cite}
\usepackage[cmex10]{amsmath}
\usepackage{amssymb}
\usepackage{makecell}
\usepackage{boldline}
\usepackage{adjustbox}
\usepackage{caption}
\usepackage{multicol}
\usepackage{multirow}
\usepackage{tikz}
\usetikzlibrary{shapes,arrows,fit,positioning,shadows,calc}
\usetikzlibrary{plotmarks}
\usetikzlibrary{decorations.pathreplacing}
\usetikzlibrary{patterns}
\usetikzlibrary{automata}
\usepackage{xcolor}
\usepackage{pgfplots}
\pgfplotsset{compat=newest}

\usepackage{hhline}

\ifCLASSINFOpdf
\else
\fi
\usepackage[cmintegrals]{newtxmath}
\begin{document}

\author{\IEEEauthorblockN{Rodrigo R.\ M.\ de Alencar,
 Lukas T.\ N.\ Landau and Rodrigo C.\ de Lamare }
\IEEEauthorblockA{Centro de Estudos em Telecomunica\c{c}\~{o}es, Pontif\'{i}cia Universidade Cat\'{o}lica do Rio de Janeiro\\ Rio de Janeiro CEP 22453-900, Brazil, Email: \{alencar, lukas.landau, delamare\}@cetuc.puc-rio.br}
}

\title{Continuous Phase Modulation \\ With 1-Bit Quantization and Oversampling \\ Using Iterative Detection and Decoding}

\maketitle
\begin{abstract}
A channel with continuous phase modulation and 1-bit ADC with oversampling is considered. 
Due to oversampling, higher-order modulations yield a higher achievable rate and this work presents a method to approach this with sophisticated coding.
A tailored bit mapping is proposed where two bits can be detected reliably at high SNR and only the third bit suffers from an error floor. With this, a convolutional code is only applied to the third bit.
Our experiments with a turbo receiver show that the proposed method corresponds to a lower BER in comparison to the case of conventional channel coding.
\end{abstract}


%
\IEEEpeerreviewmaketitle

\section{Introduction}

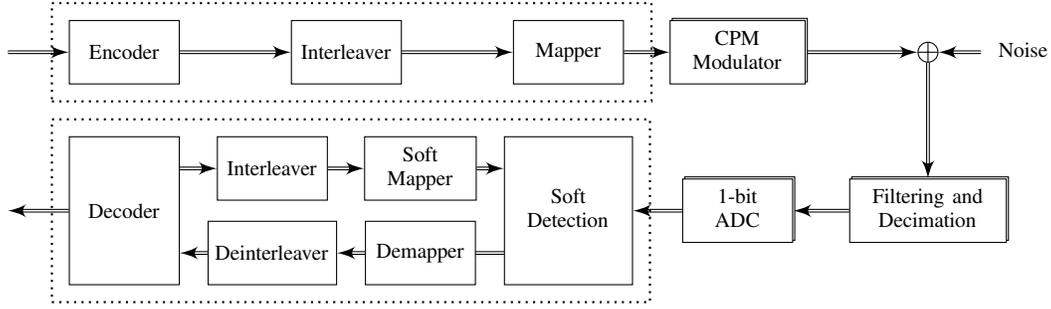
\begin{figure*}[t]
\begin{center}
\captionsetup{font=footnotesize}
\usetikzlibrary{shapes,arrows,fit,positioning,shadows,calc}
\usetikzlibrary{plotmarks}
\usetikzlibrary{decorations.pathreplacing}
\usetikzlibrary{patterns}

\tikzstyle{block} = [draw, fill=white, rectangle,minimum height=3em, minimum width=5.2em]	
\tikzstyle{block_rot} = [draw, fill=white, rectangle,minimum height=3.8em, minimum width=2em]
\tikzstyle{sum} = [draw, fill=white, circle, node distance=1em,path picture={\draw[black](path picture bounding box.south) -- (path picture bounding box.north) (path picture bounding box.west) -- (path picture bounding box.east);}]
\tikzstyle{coord} = [coordinate]

\tikzstyle{state}=[shape=circle,draw=blueTUD50,fill=blueTUD10]	
\tikzstyle{lightedge}=[<-,dotted]
\tikzstyle{mainstate}=[state,thick]
\tikzstyle{mainedge}=[<-,thick]

\tikzstyle{symbol}=[shape=circle,draw=blueTUD50,fill=blueTUD10,minimum width=1em,scale=0.6]	
\tikzstyle{sample1}=[shape=circle,draw,scale=0.3]	
\tikzstyle{sample2}=[shape=circle,draw,densely dashed,scale=0.3]	
\tikzstyle{interleave}=[shape=circle,draw,fill,minimum width=1em,scale=0.6]	

\tikzstyle{register} = [draw, fill=white, rectangle,minimum height=3em, minimum width=3em]	
\tikzstyle{mod2} = [draw, fill=white, circle, label={mod 2}, node distance=1em,path picture={\draw[black](path picture bounding box.south) -- (path picture bounding box.north) (path picture bounding box.west) -- (path picture bounding box.east);}]

\tikzset{multiple/.style={copy shadow={shadow xshift=0.1em,shadow yshift=0.1em}, draw=black,fill=white, rectangle,minimum height=3em, minimum width=5.2em},
multiple_rot/.style={copy shadow={shadow xshift=0.1em,shadow yshift=0.1em}, draw=black, fill=white, rectangle,minimum height=3.8em, minimum width=2em}}

\tikzset{>=latex,every picture/.style={font issue=\footnotesize},
         font issue/.style={execute at begin picture={#1\selectfont}}
        }

\pgfdeclarelayer{bg}    
\pgfsetlayers{bg,main}  

\begin{tikzpicture}[auto, node distance=4em,>=latex']
    
    \node[coord,  node distance=0em, align=center](source){};
	
	\node [block, right of=source, node distance=5.5em] (convenc) {Encoder};
	\node [block, right of=convenc, node distance=10.5em] (intlvr) {Interleaver};
	\node [block, right of=intlvr, node distance=10.5em] (map) {Mapper};
    \node [multiple, right of=map, node distance=8em] (ht) {\begin{tabular}{c} CPM  \\  Modulator \end{tabular}};

     \node [sum,right of=ht,node distance=9em] (Sum1) {};
	
	 \node [multiple, below of=Sum1, node distance=7.5em] (gtdec) {\begin{tabular}{c} Filtering and \\ Decimation
     \end{tabular}};
		
     \node [coord, right of=Sum1,node distance=2.5em] (noise) {};
     \node [coord, label={Noise},yshift=-0.6em, right of=noise, node distance=2em] (noise_label) {};
     \node [multiple, left of=gtdec, node distance=9em] (adc) {\begin{tabular}{c} 1-bit  \\ ADC
     \end{tabular}};
     \node [block, left of=adc, node distance=8em, minimum height=7em] (bcjr) {\begin{tabular}{c} Soft  \\ Detection
     \end{tabular}};
     \node [coord, below of=bcjr, node distance=2em] (bcjr_down) {};
     \node [coord, above of=bcjr, node distance=2em] (bcjr_up) {};
     \node [block, left of=bcjr_down, node distance=7em] (demap) {Demapper};
     \node [block, left of=demap, node distance=7em] (deintlvr) {Deinterleaver};
     \node [coord, left of=deintlvr, node distance=7em] (viterbi_down) {};
     \node [block, above of=viterbi_down, node distance=2em, minimum height=7em] (viterbi) {Decoder};
     \node [coord, above of=viterbi, node distance=2em] (viterbi_up) {};
     \node [block, right of=viterbi_up, node distance=7em] (intlvr2) {Interleaver};
     \node [block, right of=intlvr2, node distance=7em] (softmap) {\begin{tabular}{c} Soft  \\ Mapper
     \end{tabular}};
     \node [coord,node distance=5.5em, left of=viterbi] (output) {};
%
 		
	
     
     \begin{pgfonlayer}{bg}    
     
         \draw [->,double] (source) -- node {} (convenc);
         \draw [->,double] (convenc) -- node {} (intlvr);
         \draw [->,double] (intlvr) -- node {} (map);
         \draw [->,double] (map) -- node {} (ht);
         \draw [->,double] (ht) -- node {}(Sum1);
         \draw [->,double] (noise) -- node {}(Sum1);
         \draw [->,double] (Sum1) -- node {}(gtdec);
         \draw [->,double] (gtdec) -- node {}(adc);
         \draw [->,double] (adc) -- node {}(bcjr);
         \draw [->,double] (bcjr_down) -- node {}(deintlvr);
         \draw [->,double] (deintlvr) -- node {}(viterbi.east|-viterbi_down);
         \draw [->,double] (viterbi) -- node {}(output);
         \draw [->,double] (viterbi_up) -- node {}(intlvr2);
         \draw [->,double] (intlvr2) -- node {}(softmap);
         \draw [->,double] (softmap) -- node {}(bcjr.west|-bcjr_up);
         
         \draw[thick,dotted] ($(convenc.north west)+(-0.8em,0.8em)$) rectangle ($(map.south east)+(1.3em,-0.8em)$);
         \draw[thick,dotted] ($(viterbi.north west)+(-0.8em,0.8em)$) rectangle ($(bcjr.south east)+(0.8em,-0.8em)$);
     
     \end{pgfonlayer}
\end{tikzpicture}
\caption{ Extended discrete system model, 1-bit quantization, oversampling at the receiver and coding blocks with an iterative decoding strategy} 
\label{fig:discrete_system}       
\end{center}
\vspace{-0.5em}
\end{figure*}

\IEEEPARstart{T}{he} energy consumption of an
 analog-to-digital converter (ADC) scales exponentially with its resolution in amplitude. 
  Hence, in this study the receiver has only sign information about the received signal. Oversampling with respect to the temporal duration of a transmit symbol is used to compensate for the loss in terms of achievable rate. 
This receive processing approach is considered in combination with continuous phase modulation (CPM) as studied before in \cite{Landau_CPM_2018,Bender_SPAWC2019}. CPM signals are spectral efficient, having smooth phase transitions and constant envelope \cite{Anderson_1986,Sundberg_1986}, which allows for energy efficient power amplifiers. The information is implicitly conveyed in phase transitions, which makes the use of oversampling promising in the presence of coarse quantization at the receiver.

The approach presented in \cite{Landau_CL2017} shows that the achievable rate of the strictly bandlimited channel can be lower-bounded by using an auxiliary channel law.
More recently, a lower bound on the achievable rate for CPM with 1-bit quantization and oversampling at the receiver is computed in \cite{Landau_CPM_2018}, where it is shown how oversampling increases the achievable rate.

In this study we consider the design and analysis of CPM schemes with 1-bit quantization and oversampling at the receiver, employing convolutional codes in scenarios with higher modulation order. For such cases, based on the achievable rates computed in \cite{Landau_CPM_2018} channel coding is essential for establishing reliable communications. In this context, our main contribution is the extension of the discrete system model for CPM signals received with 1-bit quantization and oversampling, presented in \cite{Landau_CPM_2018}, for a sophisticated coding and decoding scheme. 
The proposed channel coding method implies the processing of soft information using a BCJR algorithm which is part of an iterative decoding strategy.
Finally, a sophisticated channel coding scheme is proposed which exploits the special properties of the channel with oversampling.

The rest of the paper is organized as follows. In Section~\ref{sec:system_model} the discrete system model is presented, taking into account the ADC and channel coding with iterative detection and decoding. In Section~\ref{sec:cpm_1bit}, the discrete time description of the quantized CPM signal is presented. In Section~\ref{sec:soft_detection} discusses the soft detection process. The iterative decoding strategy is described in Section~\ref{sec:channel_coding}. The proposed methods to enhance the performance in terms of BER are presented Section~\ref{sec:proposed_enhancements}. Numerical results are shown in Section~\ref{sec:numerical_results} and Section~\ref{sec:conclusion} gives the conclusions.

Notation:
Bold symbols denote vectors, namely oversampling vectors, e.g., $\boldsymbol{y}_k$ is a column vector with $M$ entries, where $k$ indicates the $k$th symbol in time or rather its corresponding time interval. Bold capital symbols denote matrizes.
Sequences are denoted with $x^n= [x_1,\ldots,x_n]^T$. Likewise, sequences of vectors are written as $\boldsymbol{y}^n= [\boldsymbol{y}_1^T,\ldots,\boldsymbol{y}_n^T]^T$.
A segment of a sequence is denoted as $x^{k}_{k-L}=[ x_{k-L}, \ldots,  x_k   ]^T$. 

\section{The System Model}
\label{sec:system_model}

The system model considered in this study is an extension of the discrete system model proposed in \cite{Landau_CPM_2018} for CPM with 1-bit quantization and oversampling at the receiver. Fig.~\ref{fig:discrete_system} illustrates the extension in terms of the additional coding blocks. The purpose of this extension is to design a system for
reliable communication 
by considering sophisticated forward error correction. 

On the transmit path, the channel encoder receives information bits and generates an encoded message adding redundant information. The encoded message is interleaved to protect the coded information against burst errors. Then, the interleaved bits are grouped according to the modulation order and mapped to CPM symbols. After that, a signal is generated by a CPM modulator and noise is applied.

On the receive path the signal is filtered and quantized by a 1-bit ADC. The quantized data are then processed by an iterative detection and decoding scheme.
First the binary samples are processed by a soft detection algorithm. Then the soft information is converted to bit oriented log-likelihood ratios, which are deinterleaved subsequently. Finally, the soft information is given to the channel decoder, which returns extrinsic soft information to the detection algorithm via an interleaver and a soft mapper. In the sequel the individual blocks are described in detailed.

\subsection{CPM with 1-Bit Quantization}
\label{sec:cpm_1bit}

The CPM signal in the passband with carrier frequency $f_0$ \cite{Anderson_1986} is described by  
\begin{align}
s(t)  = \mathrm{Re}   \left\{    \sqrt{ \frac{ 2  E_{\textrm{s}}}{T_{\textrm{s}}} }         e^{ j \left( 2 \pi f_0 t  + \phi\left(t\right)\right)    }         \right\}   \text{.}
\label{eq:cpm:bp}
\end{align}
The phase term is given by
\begin{align}
\phi\left(t\right) = 2 \pi h  \sum_{k=0}^{\infty}  \alpha_{k}   f(t-k T_{\textrm{s}})  +\varphi_0   \text{,} 
\label{eq:cpm:phase_term}
\end{align}
where $T_{\textrm{s}}$ denotes the symbol duration, $h=\frac{K_{\textrm{cpm}}}{P_{\textrm{cpm}}}$ is the modulation index, $f\left(\cdot\right)$ is the phase response, $\varphi_0$ is a phase-offset and $\alpha_k$ represents the $k^{\textrm{th}}$ transmit symbol with symbol energy $E_{\textrm{s}}$.
The parameters $K_{\textrm{cpm}}$ and $P_{\textrm{cpm}}$ must be relative prime positive integers in order to obtain a finite number of phase states.
For an even modulation order $M_{\textrm{cpm}}$, the symbol alphabet can be described by $\alpha_k \in \left\{  \pm 1, \pm 3,\ldots  ,\pm   (M_{\textrm{cpm}}-1)   \right\}$.
The phase response function shapes the phase transition between the phase states. It fulfills the following condition
\begin{align}
f(\tau)=\begin{cases}
  0,  & \text{ if } \tau\leq 0{ ,}  \notag  \\  
  \frac{1}{2}, & \text{ if }   \tau >  L_{\textrm{cpm}} T_{\textrm{s}}  \text{,}
\end{cases}   
\end{align}
where $L_{\textrm{cpm}}$ is the depth of the memory in terms of transmit symbols. The phase response corresponds to the integration over the frequency pulse $g_{f}\left(\cdot\right)$.


Generally, the corresponding phase trellis of \eqref{eq:cpm:phase_term} is time variant, which means that the possible phase states are time-dependent. Because of that, the number of states can be larger than $M_{\textrm{cpm}}$, e.g., when $M_{\textrm{cpm}}=2$ and $h=\frac{1}{2}$, there are at least four trellis states in total and even more depending on the memory of the channel. In order to avoid the time-dependency, a time invariant trellis is constructed by tilting the phase according to the decomposition approach in \cite{Rimoldi_1988}. The tilt corresponds to a frequency offset applied to the CPM signal, i.e., the phase term becomes $\psi(t) = \phi(t)  + 2 \pi  \Delta f t $,
where $\Delta f = h (M_{\textrm{cpm}}   -1)/2  T_{\textrm{s}} $.
Taking into account the tilted trellis, a different symbol notation $x_k=  (\alpha_k  +  M_{\textrm{cpm}} - 1) / 2  $ can be considered, which then corresponds to the symbol alphabet $\mathcal{X} = \left\{  0, 1,\ldots  ,  M_{\textrm{cpm}}-1   \right\} $. The tilted CPM phase $\psi(t)$ within one symbol interval with duration $T_{\textrm{s}}$, letting $t=\tau+kT_{\textrm{s}}$, can be fully described by the state definition $\tilde{s}_{k} = \left[  \beta_{k-L_{\textrm{cpm}}} , x^{k}_{k-L_{\textrm{cpm}}+1}   \right]$ in terms of 
\begin{align}
\psi (\tau  +  k T_{\textrm{s}} ) =  &   \frac{2 \pi}{P_{\textrm{cpm}}} \beta_{k-L_{\textrm{cpm}}} \label{eq:cpm:Rimoldi}\\
& + 2 \pi h \sum_{l=0}^{L_{\textrm{cpm}}-1} \left( 2 x_{k-l} - M_{\textrm{cpm}} + 1\right) f(\tau+l T_{\textrm{s}}) \notag \\   
& + \pi h \left( M_{\textrm{cpm}} -1 \right) \left( \frac{\tau}{T_{\textrm{s}}} + L_{\textrm{cpm}} - 1 \right)  + \varphi_0 \text{,} \notag
\end{align}
where the absolute phase state
$\beta_{k-L_{\textrm{cpm}}}$ can be reduced to
\begin{align}
\beta_{k-L_{\textrm{cpm}}} = \left( K_{\textrm{cpm}}   \sum_{l=0}^{k-L_{\textrm{cpm}}}   x_l    \right) \bmod P_{\textrm{cpm}} \text{,}
\label{eq:cpm:beta}
\end{align}
which is related to the $2\pi$-wrapped accumulated phase contributions of the input symbols that are prior to the CPM memory. Note that $\tilde{s}_{k}$ is the appropriate state description for modeling one symbol duration of the signal. To model the signal as result of state transitions, while taking the receive filter into account, another state description, namely $s_k$ will be introduced later. 
For the illustrated examples $\varphi_0= \pi h $ is considered. 
In the sequel a discrete time description is considered which implies that the CPM phase is represented in a vector notation. The corresponding tilted CPM phase $\psi (\tau  +  k T_{\textrm{s}} )$ for one symbol interval, i.e.,  $0 \leq \tau < T_{\textrm{s}}$, is then discretized into $MD$ samples, which composes the vector denoted by
$\boldsymbol{\psi}_{k}(\tilde{s}_{k})  =  [\psi(\frac{T_{\textrm{s}}}{MD}(kMD+1)),\psi(\frac{T_{\textrm{s}}}{MD}(kMD+2)),\dots,\psi(T_{\textrm{s}}(k+1))]^T$, where $M$ is the oversampling factor, and $D$ is a higher resolution multiplier.


The receive filter $g(t)$ has an impulse response of length $T_g$. In the discrete model for expressing a subsequence of $(N +1)$ oversampling output symbols it is represented in a matrix form with $\boldsymbol{G}$, as a $MD(N+1) \times MD(L_g+N+1)$ Toepliz matrix, as  described in equation (17) in \cite{Landau_CPM_2018}, whose first row is $[\boldsymbol{g}^T,\boldsymbol{0}_{MD(N+1)}^T]$, where $\boldsymbol{g}^T=[g( L_g T_{\textrm{s}} ), g( \frac{T_{\textrm{s}}}{MD} (L_gMD-1)),\dots,g(\frac{T_{\textrm{s}}}{MD})]$.
A higher sampling grid in the waveform signal, in the noise generation and in the filtering is adopted to adequately model the aliasing effect. This receive filtering yields an increase of memory in the system by $L_g$ symbols, where $(L_{g}-1) T_{\textrm{s}} < T_g \leq L_{g} T_{\textrm{s}}$.

The filtered samples are decimated to the vector $\boldsymbol{z}^{k}_{k-N}$ according to the oversampling factor $M$, by multiplication with the $D$-fold decimation matrix $\boldsymbol{D}$, as described in equation (16) in \cite{Landau_CPM_2018}, with dimensions $M(N+1) \times MD(N+1)$. Then, the result $\boldsymbol{z}^{k}_{k-N}$ is 1-bit quantized to the vector $\boldsymbol{y}^{k}_{k-N}$. These operations can be represented by the following equations
\begin{align}
\boldsymbol{y}^{k}_{k-N} = Q \left( \boldsymbol{z}^{k}_{k-N}  \right) =  Q  \left(  \boldsymbol{D} \  \boldsymbol{G} \left[ \sqrt{  \frac{E_{\textrm{s}}}{T_{\textrm{s}}} }   e^{\boldsymbol{\psi}^{k}_{k-N-L_g}}    +  \boldsymbol{n}^{k}_{k-N-L_g }  \right] \right)   \text{,} \notag
\end{align}
where $Q(\cdot)$ denotes the quantization operator. The quantization of $\boldsymbol{z}_k$ is described by ${y}_{k,m} =\mathrm{sgn}(\mathrm{Re}\left\{{z}_{k,m}\right\})+j \mathrm{sgn}(\mathrm{Im}\left\{{z}_{k,m}\right\})$, where $m$ denotes the oversampling index which runs from $1$ to $M$ and ${y}_{k,m} \in \left\{ 1+j,1-j,-1+j,-1-j \right\}$. The vector $\boldsymbol{n}^{k}_{k-N-L_g}$ contains complex zero-mean white Gaussian noise samples with variance $\sigma_n^2=N_0$.


\subsection{Soft Detection}
\label{sec:soft_detection}

The MAP decision metric for each bit is given by the A Posteriori Probability (APP) given the received sequence $\boldsymbol{y}^n$, which, for the considered system, can be  approximately computed via a BCJR algorithm \cite{BCJR_1974} based on an auxiliary channel law described in \cite{Landau_CPM_2018} for computing a lower bound on the achievable rate.
Depending on the receive filter, noise samples are correlated which then implies dependency on previous channel outputs, such that the channel law has the form $P(\boldsymbol{y}_k \vert \boldsymbol{y}^{k-1}, x^n )$.
In this case the consideration of an auxiliary channel law $W(\cdot)$ is required, which reads as
\begin{align}
W(\boldsymbol{y}_k \vert \boldsymbol{y}^{k-1}, x^n ) & =
P(\boldsymbol{y}_k \vert \boldsymbol{y}^{k-1}_{k-N}, \beta_{k-L} , x^{k}_{k-L+1}) \notag \\
& =\frac{
P( \boldsymbol{y}^k_{k-N}   \vert \beta_{k-L}, x^{k}_{k-L+1})}{P(  \boldsymbol{y}^{k-1}_{k-N} \vert \beta_{k-L}, x^{k-1}_{k-L+1})}
\text{,}
\label{eq:aux_channel}
\end{align}
where the dependency on $N$ previous channel outputs is taken into account with $L=L_{\textrm{cpm}}+L_g+N$ being the total memory of the system.
With this, the BCJR algorithm relies on an extended state representation denoted by
\begin{align}
s_k=\begin{cases}
  [ \beta_{k-L+1} { ,} x^{k}_{k-L+2}  ] { ,}  & \text{ if } L> 1{ ,}   \\  
  [ \beta_{k} ] , & \text{ if }   L=1  \text{.}
\end{cases}
\label{eq:states_CPM}
\end{align}
Based on the state notation
in \eqref{eq:states_CPM} the probabilities for the channel law can be cast as
$P\left(   \boldsymbol{y}^{k}_{k-N} \vert s_k, s_{k-1} \right)$ and
$P\left(   \boldsymbol{y}^{k-1}_{k-N} \vert s_{k-1} \right)$.
The auxiliary channel law probabilities \eqref{eq:aux_channel} 
involve a multivariate Gaussian integration in terms of
\begin{align}
 P\left(   \boldsymbol{y}^{k}_{k-N} \vert s_k, s_{k-1} \right) = 
 \int\limits_{ \boldsymbol{z}_{k-N}^{k} \in  \mathbb{Y}_{k-N}^{k}}  p( \boldsymbol{z}_{k-N}^{k}  \vert s_k, s_{k-1}    )  d \boldsymbol{z}_{k-N}^{k}  \text{,}
\label{eq:multivariate_integration}
\end{align}
where $\boldsymbol{z}_{k-N}^{k}$ is a complex Gaussian random vector that describes the input of the ADC, with a mean vector defined by $\boldsymbol{\mu}_{x} = \boldsymbol{D}\  \boldsymbol{G} \left[ \sqrt{  \frac{E_{\textrm{s}}}{T_{\textrm{s}}} }   e^{\boldsymbol{\psi}^{k}_{k-N-L_g}} \right]$, and covariance matrix $\boldsymbol{R}=\sigma_n^2 \boldsymbol{D}\boldsymbol{G}\boldsymbol{G}^H\boldsymbol{D}^T$, with $\boldsymbol{D}$ and $\boldsymbol{G}$ as introduced before.
 The integration interval is expressed in terms of the quantization region $ \mathbb{Y}_{k-N}^{k}$
that belongs to the channel output symbol $\boldsymbol{y}^{k}_{k-N}$.
After rewriting \eqref{eq:multivariate_integration} as a real valued multivariate Gaussian integration, how it is done in equation (21) in \cite{Landau_CPM_2018}, the algorithm in \cite{Genz_1992} can be applied.

Finally the BCJR algorithm provides the
probabilities $P_{\textrm{aux}} \left( s_{k-1},s_{k} \vert \boldsymbol{y}^n \right)$
which are then used for computing the bit APPs, with an approach similar to the one presented in \cite{Tuchler_2011}

\begin{align}
    P\left(b_{k,i}=b\vert\boldsymbol{y}^n\right) = \sum_{\substack{\forall s_{k-1},s_{k} \supseteq x_{k} \\ \text{such that} \\ {\rm \mathrm{map}}^{-1}(x_k,i)=b}} P_{\textrm{aux}}(s_k,s_{k-1}\vert \boldsymbol{y}^n) \text{,}
\end{align}
 where $b_{k,i}={\rm \mathrm{map}}^{-1}(x_k,i)$ denotes the $i^{th}$ bit (MSB first) of the $k^{th}$ demapped symbol with $i \in   \left\{1, 2, \ldots  ,  \log_2(M_{\textrm{cpm}})\right\}$. With the values of $P\left(b_{k,i}=0\vert\boldsymbol{y}^n\right)$ and $P\left(b_{k,i}=1\vert\boldsymbol{y}^n\right)$, the evaluation of the log-likelihood ratios is done as follows

\begin{align}
    L(b_{k,i}) = \ln\left(\frac{P(b_{k,i}=0\vert\boldsymbol{y}^n)}{P(b_{k,i}=1\vert\boldsymbol{y}^n)}\right)
\end{align}

The log-likelihood ratio sequence is deinterleaved, according to the permutation adopted. The resultant sequence represents the detected soft information used as input for the channel decoder as represented in Fig.~\ref{fig:discrete_system}.

\subsection{Channel Coding and Iterative Decoding}
\label{sec:channel_coding}

For the proposed CPM system,
the lower bound on on the achievable rate presented in 
\cite{Landau_CPM_2018}
serves as the base information for choosing the code rate for channel coding.
The result of the achievable rate computation in \cite{Landau_CPM_2018} shows that the scenarios with higher modulation order, e.g., $M_{\textrm{cpm}}=8$ and $M=3$, require channel coding in order to provide communication with low probability of error, because the achievable rate at high SNR is significantly lower than the input entropy. With this, a coding scheme must satisfy the following inequality
\begin{align}
    R \cdot \log_{2}(M_{\textrm{cpm}}) \leq I_{M_{\textrm{cpm}}} \text{  [bpcu],}
\label{eq:code:condition}
\end{align}
where $R$ denotes the code rate of the channel code and $I_{M_{\textrm{cpm}}}$ denotes the achievable rate conditioned on the corresponding CPM modulation scheme. Note that $\log_{2}(M_{\textrm{cpm}})$ is the maximum entropy rate of the input, which is an upper bound of the achievable rate.

The extended system in Fig.~\ref{fig:discrete_system} shows that an interleaver block is applied. The purpose of it is to reduce the impact of burst errors that might occur due to the fact that channels are not memoryless. Those burst of errors are harmfull for the performance when using convolutional codes. The Interleaver shuffles the bits, distributing the errors along the block of bits, improving the performance of the system.

The convolutional codes can also be described by a state machine. From that a trellis structure can be derived, which is the base description of the code used in the decoding procedure, which can be implemented using the Viterbi Algorithm (ML decoder) or using the BCJR (MAP decoder). This work uses an iterative decoding procedure that consist of the exchange of soft information between the detector and the channel decoder. Initially, the state transition probabilities is considered to be uniformly distributed, i.e., $P(s_k\vert s_{k-1})=1/M_{\textrm{cpm}}$, such assumption is suboptimal, but it is possible take into account that state transitions can have different probabilities by feeding back updated extrinsic soft information about the code bits.

This extrinsic information, can be computed during the channel decoding procedure by means of employing a second BCJR algorithm. It can then be interleaved and soft mapped to new values of state transition probabilities $P\left(s_k\vert s_{k-1}\right)$. This soft mapping can be described by

\begin{align}
    P\left(s_k\vert s_{k-1}\right) & = \prod_{i=1}^{\log_2(M_{\textrm{cpm}})} \frac{\exp\left( - b_{k,i}\cdot L_{ext}\left( b_{k,i} \right)\right)}{1+\exp\left( -L_{ext}\left( b_{k,i} \right)\right)} \\
    x_k & = \mathrm{map}\left(b_{k,1},\dots,b_{k,\log_2(M_{\textrm{cpm}})}\right) \notag
\label{eq:code:softmap}
\end{align}
where $x_k$ is the output symbol from the state transition of $s_{k-1}$ to $s_k$, $[b_{k,1},\dots,b_{k,\log_2(M_{\textrm{cpm}})}]$ is the bit sequence which such symbol is mapped to, and $L_{ext}(b)$ represents the interleaved soft information fed back by the channel decoder of the bit $b$. The probabilities $P\left(s_k\vert s_{k-1}\right)$ can be used to perform the soft detection step again by updating the transition probabilities in the BCJR algorithm. All this procedure is described in \cite{Tuchler_2011} as a turbo equalization technique.

\section{Proposed Enhancements}
\label{sec:proposed_enhancements}

In the present study different coding and mapping strategies are considered and illustrated it for the example of $M_{\textrm{cpm}}=8$.
\begin{figure}[t!]
	\centering
	\begin{tabular}{ c c }
	\multirow{2}{*}[6.5em]{
	    \scalebox{.7}{\definecolor{mycolor1}{rgb}{0.00000,0.44700,0.74100}%
\definecolor{mycolor2}{rgb}{0.85000,0.32500,0.09800}%
\definecolor{mycolor3}{rgb}{0.92900,0.69400,0.12500}%
\begin{tikzpicture}

\begin{axis}[%
width=15em,
height=23em,
scale only axis,
grid style={color=gray!15},
xmin=-5,
xmax=41,
xlabel={SNR [dB]},
ymode=log,
ymin=0.0001,
ymax=1,
yminorticks=true,
ylabel={BER},
xmajorgrids,
ymajorgrids,
yminorgrids,
legend style={at={(0,0)},anchor=south west,draw=black,fill=white,legend cell align=left,font=\small}
]
\addplot [color=mycolor1, mark=square*, mark options={solid, mycolor1}]
  table[row sep=crcr]{%
-4.37014015627852	0.429939393939394\\
0.629859843721482	0.3215\\
5.62985984372148	0.167941712204007\\
10.6298598437215	0.0810810810810811\\
15.6298598437215	0.0472231404958678\\
20.6298598437215	0.029355474673946\\
25.6298598437215	0.0214737808951236\\
30.6298598437215	0.016988416988417\\
35.6298598437215	0.0153726472374013\\
40.6298598437215	0.0151267162391734\\
};
\addlegendentry{$\text{1}^{\text{st}}\text{ bit sub-channel}$}

\addplot [color=mycolor2, mark=square*, mark options={solid, mycolor2}]
  table[row sep=crcr]{%
-4.37014015627852	0.450575757575758\\
0.629859843721482	0.372866666666667\\
5.62985984372148	0.256675774134791\\
10.6298598437215	0.165635635635636\\
15.6298598437215	0.12932782369146\\
20.6298598437215	0.101789505611162\\
25.6298598437215	0.0807105878423514\\
30.6298598437215	0.0682289943004229\\
35.6298598437215	0.0647100789313904\\
40.6298598437215	0.0654293034784304\\
};
\addlegendentry{$\text{2}^{\text{nd}}\text{ bit sub-channel}$}

\addplot [color=mycolor3, mark=square*, mark options={solid, mycolor3}]
  table[row sep=crcr]{%
-4.37014015627852	0.49230303030303\\
0.629859843721482	0.475233333333333\\
5.62985984372148	0.413688524590164\\
10.6298598437215	0.309249249249249\\
15.6298598437215	0.229845730027548\\
20.6298598437215	0.183522899605702\\
25.6298598437215	0.155460086840347\\
30.6298598437215	0.134531163816878\\
35.6298598437215	0.124832776765837\\
40.6298598437215	0.125255242710335\\
};
\addlegendentry{$\text{3}^{\text{rd}}\text{ bit sub-channel}$}

\end{axis}
\end{tikzpicture}%
}
	}
	&
    \subfloat["Even" State]{
        \scalebox{.6}{\tikzset{every picture/.style={line width=0.75pt}} 

\begin{tikzpicture}[x=0.75pt,y=0.75pt,yscale=-0.7,xscale=0.7]
\clip (66,0) rectangle + (260,240);

\draw [color={rgb, 255:red, 109; green, 109; blue, 109 }  ,draw opacity=1 ]   (173.13,250) -- (173.69,12.76) ;
\draw [shift={(173.7,10.76)}, rotate = 450.14] [color={rgb, 255:red, 109; green, 109; blue, 109 }  ,draw opacity=1 ][line width=0.75]    (10.93,-3.29) .. controls (6.95,-1.4) and (3.31,-0.3) .. (0,0) .. controls (3.31,0.3) and (6.95,1.4) .. (10.93,3.29)   ;

\draw [color={rgb, 255:red, 109; green, 109; blue, 109 }  ,draw opacity=1 ]   (52,130.38) -- (292.83,130.38) ;
\draw [shift={(294.83,130.38)}, rotate = 180] [color={rgb, 255:red, 109; green, 109; blue, 109 }  ,draw opacity=1 ][line width=0.75]    (10.93,-3.29) .. controls (6.95,-1.4) and (3.31,-0.3) .. (0,0) .. controls (3.31,0.3) and (6.95,1.4) .. (10.93,3.29)   ;

\draw   (100.96,131.09) .. controls (100.67,92.11) and (132.49,60.28) .. (172.02,60) .. controls (211.56,59.72) and (243.84,91.11) .. (244.13,130.09) .. controls (244.41,169.08) and (212.6,200.91) .. (173.06,201.19) .. controls (133.52,201.46) and (101.24,170.08) .. (100.96,131.09) -- cycle ;
\draw  [fill={rgb, 255:red, 0; green, 0; blue, 0 }  ,fill opacity=1 ] (143.64,65.31) .. controls (143.64,64.1) and (144.64,63.12) .. (145.86,63.12) .. controls (147.09,63.12) and (148.08,64.1) .. (148.08,65.31) .. controls (148.08,66.52) and (147.09,67.5) .. (145.86,67.5) .. controls (144.64,67.5) and (143.64,66.52) .. (143.64,65.31) -- cycle ;
\draw  [fill={rgb, 255:red, 0; green, 0; blue, 0 }  ,fill opacity=1 ] (197.61,65.31) .. controls (197.61,64.1) and (198.61,63.12) .. (199.83,63.12) .. controls (201.06,63.12) and (202.05,64.1) .. (202.05,65.31) .. controls (202.05,66.52) and (201.06,67.5) .. (199.83,67.5) .. controls (198.61,67.5) and (197.61,66.52) .. (197.61,65.31) -- cycle ;
\draw  [fill={rgb, 255:red, 0; green, 0; blue, 0 }  ,fill opacity=1 ] (103.6,104.8) .. controls (103.6,103.59) and (104.59,102.61) .. (105.82,102.61) .. controls (107.04,102.61) and (108.04,103.59) .. (108.04,104.8) .. controls (108.04,106.01) and (107.04,106.99) .. (105.82,106.99) .. controls (104.59,106.99) and (103.6,106.01) .. (103.6,104.8) -- cycle ;
\draw  [fill={rgb, 255:red, 0; green, 0; blue, 0 }  ,fill opacity=1 ] (104.47,158.02) .. controls (104.47,156.81) and (105.46,155.83) .. (106.69,155.83) .. controls (107.91,155.83) and (108.91,156.81) .. (108.91,158.02) .. controls (108.91,159.23) and (107.91,160.21) .. (106.69,160.21) .. controls (105.46,160.21) and (104.47,159.23) .. (104.47,158.02) -- cycle ;
\draw  [fill={rgb, 255:red, 0; green, 0; blue, 0 }  ,fill opacity=1 ] (235.91,103.94) .. controls (235.91,102.73) and (236.91,101.75) .. (238.13,101.75) .. controls (239.36,101.75) and (240.35,102.73) .. (240.35,103.94) .. controls (240.35,105.15) and (239.36,106.13) .. (238.13,106.13) .. controls (236.91,106.13) and (235.91,105.15) .. (235.91,103.94) -- cycle ;
\draw  [fill={rgb, 255:red, 0; green, 0; blue, 0 }  ,fill opacity=1 ] (144.51,196.65) .. controls (144.51,195.44) and (145.51,194.46) .. (146.73,194.46) .. controls (147.96,194.46) and (148.95,195.44) .. (148.95,196.65) .. controls (148.95,197.86) and (147.96,198.84) .. (146.73,198.84) .. controls (145.51,198.84) and (144.51,197.86) .. (144.51,196.65) -- cycle ;
\draw  [fill={rgb, 255:red, 0; green, 0; blue, 0 }  ,fill opacity=1 ] (198.48,194.93) .. controls (198.48,193.72) and (199.48,192.74) .. (200.7,192.74) .. controls (201.93,192.74) and (202.92,193.72) .. (202.92,194.93) .. controls (202.92,196.14) and (201.93,197.12) .. (200.7,197.12) .. controls (199.48,197.12) and (198.48,196.14) .. (198.48,194.93) -- cycle ;
\draw  [fill={rgb, 255:red, 0; green, 0; blue, 0 }  ,fill opacity=1 ] (236.78,156.3) .. controls (236.78,155.09) and (237.78,154.11) .. (239,154.11) .. controls (240.23,154.11) and (241.22,155.09) .. (241.22,156.3) .. controls (241.22,157.51) and (240.23,158.49) .. (239,158.49) .. controls (237.78,158.49) and (236.78,157.51) .. (236.78,156.3) -- cycle ;
\draw [color={rgb, 255:red, 109; green, 109; blue, 109 }  ,draw opacity=1 ]   (238.13,103.94) .. controls (289.02,166.93) and (328.92,92.46) .. (239.62,103.6) ;
\draw [shift={(238.27,103.77)}, rotate = 352.45] [fill={rgb, 255:red, 109; green, 109; blue, 109 }  ,fill opacity=1 ][line width=0.75]  [draw opacity=0] (10.72,-5.15) -- (0,0) -- (10.72,5.15) -- (7.12,0) -- cycle    ;

\draw [color={rgb, 255:red, 109; green, 109; blue, 109 }  ,draw opacity=1 ]   (238.13,103.94) .. controls (247.3,75.97) and (225.75,50.67) .. (201.33,64.42) ;
\draw [shift={(199.83,65.31)}, rotate = 327.95] [fill={rgb, 255:red, 109; green, 109; blue, 109 }  ,fill opacity=1 ][line width=0.75]  [draw opacity=0] (10.72,-5.15) -- (0,0) -- (10.72,5.15) -- (7.12,0) -- cycle    ;

\draw [color={rgb, 255:red, 109; green, 109; blue, 109 }  ,draw opacity=1 ]   (238.27,103.77) .. controls (258.5,36.08) and (165.06,26.74) .. (146.65,63.59) ;
\draw [shift={(145.86,65.31)}, rotate = 292.79] [fill={rgb, 255:red, 109; green, 109; blue, 109 }  ,fill opacity=1 ][line width=0.75]  [draw opacity=0] (10.72,-5.15) -- (0,0) -- (10.72,5.15) -- (7.12,0) -- cycle    ;

\draw [color={rgb, 255:red, 109; green, 109; blue, 109 }  ,draw opacity=1 ]   (238.13,103.94) .. controls (276.89,9.77) and (85.77,16.95) .. (105.5,103.49) ;
\draw [shift={(105.82,104.8)}, rotate = 255.94] [fill={rgb, 255:red, 109; green, 109; blue, 109 }  ,fill opacity=1 ][line width=0.75]  [draw opacity=0] (10.72,-5.15) -- (0,0) -- (10.72,5.15) -- (7.12,0) -- cycle    ;

\draw [color={rgb, 255:red, 109; green, 109; blue, 109 }  ,draw opacity=1 ]   (238.13,103.94) .. controls (182.48,31.98) and (104.64,87.71) .. (106.65,156.97) ;
\draw [shift={(106.69,158.02)}, rotate = 267.68] [fill={rgb, 255:red, 109; green, 109; blue, 109 }  ,fill opacity=1 ][line width=0.75]  [draw opacity=0] (10.72,-5.15) -- (0,0) -- (10.72,5.15) -- (7.12,0) -- cycle    ;

\draw [color={rgb, 255:red, 109; green, 109; blue, 109 }  ,draw opacity=1 ]   (238.13,103.94) .. controls (166.89,29.42) and (89.84,146.32) .. (145.88,195.91) ;
\draw [shift={(146.73,196.65)}, rotate = 220.44] [fill={rgb, 255:red, 109; green, 109; blue, 109 }  ,fill opacity=1 ][line width=0.75]  [draw opacity=0] (10.72,-5.15) -- (0,0) -- (10.72,5.15) -- (7.12,0) -- cycle    ;

\draw [color={rgb, 255:red, 109; green, 109; blue, 109 }  ,draw opacity=1 ]   (238.27,103.77) .. controls (156.5,35.39) and (103.52,185.48) .. (199.25,194.8) ;
\draw [shift={(200.7,194.93)}, rotate = 184.65] [fill={rgb, 255:red, 109; green, 109; blue, 109 }  ,fill opacity=1 ][line width=0.75]  [draw opacity=0] (10.72,-5.15) -- (0,0) -- (10.72,5.15) -- (7.12,0) -- cycle    ;

\draw [color={rgb, 255:red, 109; green, 109; blue, 109 }  ,draw opacity=1 ]   (238.27,103.77) .. controls (135.71,52.48) and (140.37,213.69) .. (237.53,157.17) ;
\draw [shift={(239,156.3)}, rotate = 509.09] [fill={rgb, 255:red, 109; green, 109; blue, 109 }  ,fill opacity=1 ][line width=0.75]  [draw opacity=0] (10.72,-5.15) -- (0,0) -- (10.72,5.15) -- (7.12,0) -- cycle    ;

\draw (265.51,96.69) node  [align=left] {000};
\draw  [draw opacity=0][fill={rgb, 255:red, 255; green, 255; blue, 255 }  ,fill opacity=0.5 ]  (226.34, 64.92) circle [x radius= 21.21, y radius= 12.73]   ;
\draw (226.34,64.92) node  [align=left] {001};
\draw  [draw opacity=0][fill={rgb, 255:red, 255; green, 255; blue, 255 }  ,fill opacity=0.5 ]  (161.92, 48.61) circle [x radius= 19.8, y radius= 12.73]   ;
\draw (161.92,48.61) node  [align=left] {011};
\draw (87.06,88.96) node  [align=left] {010};
\draw (87.06,159.35) node  [align=left] {110};
\draw (129.71,205.71) node  [align=left] {111};
\draw (211.54,207.42) node  [align=left] {101};
\draw (254.19,164.5) node  [align=left] {100};
\draw (289,145.42) node [scale=0.9]  {$Re\left\{e^{j\psi ( t)}\right\}$};
\draw (220,14) node [scale=0.9]  {$Im\left\{e^{j\psi ( t)}\right\}$};

\end{tikzpicture}}
    }

	\\
	&
	\subfloat["Odd" State]{
	    \scalebox{.6}{\tikzset{every picture/.style={line width=0.75pt}} 

\begin{tikzpicture}[x=0.75pt,y=0.75pt,yscale=-0.7,xscale=0.7]
\clip (15,0) rectangle + (260,240);

\draw [color={rgb, 255:red, 109; green, 109; blue, 109 }  ,draw opacity=1 ]   (121.13,250) -- (121.69,12.76) ;
\draw [shift={(121.7,10.76)}, rotate = 450.14] [color={rgb, 255:red, 109; green, 109; blue, 109 }  ,draw opacity=1 ][line width=0.75]    (10.93,-3.29) .. controls (6.95,-1.4) and (3.31,-0.3) .. (0,0) .. controls (3.31,0.3) and (6.95,1.4) .. (10.93,3.29)   ;

\draw [color={rgb, 255:red, 109; green, 109; blue, 109 }  ,draw opacity=1 ]   (0,130.38) -- (240.83,130.38) ;
\draw [shift={(242.83,130.38)}, rotate = 180] [color={rgb, 255:red, 109; green, 109; blue, 109 }  ,draw opacity=1 ][line width=0.75]    (10.93,-3.29) .. controls (6.95,-1.4) and (3.31,-0.3) .. (0,0) .. controls (3.31,0.3) and (6.95,1.4) .. (10.93,3.29)   ;

\draw   (48.96,131.09) .. controls (48.67,92.11) and (80.49,60.28) .. (120.02,60) .. controls (159.56,59.72) and (191.84,91.11) .. (192.13,130.09) .. controls (192.41,169.08) and (160.6,200.91) .. (121.06,201.19) .. controls (81.52,201.46) and (49.24,170.08) .. (48.96,131.09) -- cycle ;
\draw  [fill={rgb, 255:red, 0; green, 0; blue, 0 }  ,fill opacity=1 ] (91.64,65.31) .. controls (91.64,64.1) and (92.64,63.12) .. (93.86,63.12) .. controls (95.09,63.12) and (96.08,64.1) .. (96.08,65.31) .. controls (96.08,66.52) and (95.09,67.5) .. (93.86,67.5) .. controls (92.64,67.5) and (91.64,66.52) .. (91.64,65.31) -- cycle ;
\draw  [fill={rgb, 255:red, 0; green, 0; blue, 0 }  ,fill opacity=1 ] (145.61,65.31) .. controls (145.61,64.1) and (146.61,63.12) .. (147.83,63.12) .. controls (149.06,63.12) and (150.05,64.1) .. (150.05,65.31) .. controls (150.05,66.52) and (149.06,67.5) .. (147.83,67.5) .. controls (146.61,67.5) and (145.61,66.52) .. (145.61,65.31) -- cycle ;
\draw  [fill={rgb, 255:red, 0; green, 0; blue, 0 }  ,fill opacity=1 ] (51.6,104.8) .. controls (51.6,103.59) and (52.59,102.61) .. (53.82,102.61) .. controls (55.04,102.61) and (56.04,103.59) .. (56.04,104.8) .. controls (56.04,106.01) and (55.04,106.99) .. (53.82,106.99) .. controls (52.59,106.99) and (51.6,106.01) .. (51.6,104.8) -- cycle ;
\draw  [fill={rgb, 255:red, 0; green, 0; blue, 0 }  ,fill opacity=1 ] (52.47,158.02) .. controls (52.47,156.81) and (53.46,155.83) .. (54.69,155.83) .. controls (55.91,155.83) and (56.91,156.81) .. (56.91,158.02) .. controls (56.91,159.23) and (55.91,160.21) .. (54.69,160.21) .. controls (53.46,160.21) and (52.47,159.23) .. (52.47,158.02) -- cycle ;
\draw  [fill={rgb, 255:red, 0; green, 0; blue, 0 }  ,fill opacity=1 ] (183.91,103.94) .. controls (183.91,102.73) and (184.91,101.75) .. (186.13,101.75) .. controls (187.36,101.75) and (188.35,102.73) .. (188.35,103.94) .. controls (188.35,105.15) and (187.36,106.13) .. (186.13,106.13) .. controls (184.91,106.13) and (183.91,105.15) .. (183.91,103.94) -- cycle ;
\draw  [fill={rgb, 255:red, 0; green, 0; blue, 0 }  ,fill opacity=1 ] (92.51,196.65) .. controls (92.51,195.44) and (93.51,194.46) .. (94.73,194.46) .. controls (95.96,194.46) and (96.95,195.44) .. (96.95,196.65) .. controls (96.95,197.86) and (95.96,198.84) .. (94.73,198.84) .. controls (93.51,198.84) and (92.51,197.86) .. (92.51,196.65) -- cycle ;
\draw  [fill={rgb, 255:red, 0; green, 0; blue, 0 }  ,fill opacity=1 ] (146.48,194.93) .. controls (146.48,193.72) and (147.48,192.74) .. (148.7,192.74) .. controls (149.93,192.74) and (150.92,193.72) .. (150.92,194.93) .. controls (150.92,196.14) and (149.93,197.12) .. (148.7,197.12) .. controls (147.48,197.12) and (146.48,196.14) .. (146.48,194.93) -- cycle ;
\draw  [fill={rgb, 255:red, 0; green, 0; blue, 0 }  ,fill opacity=1 ] (184.78,156.3) .. controls (184.78,155.09) and (185.78,154.11) .. (187,154.11) .. controls (188.23,154.11) and (189.22,155.09) .. (189.22,156.3) .. controls (189.22,157.51) and (188.23,158.49) .. (187,158.49) .. controls (185.78,158.49) and (184.78,157.51) .. (184.78,156.3) -- cycle ;
\draw [color={rgb, 255:red, 109; green, 109; blue, 109 }  ,draw opacity=1 ]   (148.29,64.81) .. controls (228.82,73.37) and (204.37,-7.5) .. (149.11,63.52) ;
\draw [shift={(148.27,64.6)}, rotate = 307.45] [fill={rgb, 255:red, 109; green, 109; blue, 109 }  ,fill opacity=1 ][line width=0.75]  [draw opacity=0] (10.72,-5.15) -- (0,0) -- (10.72,5.15) -- (7.12,0) -- cycle    ;

\draw [color={rgb, 255:red, 109; green, 109; blue, 109 }  ,draw opacity=1 ]   (148.29,64.81) .. controls (135,38.55) and (101.87,35.91) .. (94.32,62.89) ;
\draw [shift={(93.9,64.58)}, rotate = 282.95] [fill={rgb, 255:red, 109; green, 109; blue, 109 }  ,fill opacity=1 ][line width=0.75]  [draw opacity=0] (10.72,-5.15) -- (0,0) -- (10.72,5.15) -- (7.12,0) -- cycle    ;

\draw [color={rgb, 255:red, 109; green, 109; blue, 109 }  ,draw opacity=1 ]   (148.27,64.6) .. controls (114.71,2.43) and (42.03,61.9) .. (55.08,100.97) ;
\draw [shift={(55.73,102.75)}, rotate = 247.79000000000002] [fill={rgb, 255:red, 109; green, 109; blue, 109 }  ,fill opacity=1 ][line width=0.75]  [draw opacity=0] (10.72,-5.15) -- (0,0) -- (10.72,5.15) -- (7.12,0) -- cycle    ;

\draw [color={rgb, 255:red, 109; green, 109; blue, 109 }  ,draw opacity=1 ]   (148.29,64.81) .. controls (109.11,-29.18) and (-20.96,111.04) .. (54.19,158.28) ;
\draw [shift={(55.34,158.98)}, rotate = 210.94] [fill={rgb, 255:red, 109; green, 109; blue, 109 }  ,fill opacity=1 ][line width=0.75]  [draw opacity=0] (10.72,-5.15) -- (0,0) -- (10.72,5.15) -- (7.12,0) -- cycle    ;

\draw [color={rgb, 255:red, 109; green, 109; blue, 109 }  ,draw opacity=1 ]   (148.29,64.81) .. controls (58.06,53.28) and (42.42,147.73) .. (92.82,195.29) ;
\draw [shift={(93.59,196)}, rotate = 222.68] [fill={rgb, 255:red, 109; green, 109; blue, 109 }  ,fill opacity=1 ][line width=0.75]  [draw opacity=0] (10.72,-5.15) -- (0,0) -- (10.72,5.15) -- (7.12,0) -- cycle    ;

\draw [color={rgb, 255:red, 109; green, 109; blue, 109 }  ,draw opacity=1 ]   (148.29,64.81) .. controls (45.22,62.49) and (73.4,199.65) .. (148.09,195.08) ;
\draw [shift={(149.22,195)}, rotate = 535.44] [fill={rgb, 255:red, 109; green, 109; blue, 109 }  ,fill opacity=1 ][line width=0.75]  [draw opacity=0] (10.72,-5.15) -- (0,0) -- (10.72,5.15) -- (7.12,0) -- cycle    ;

\draw [color={rgb, 255:red, 109; green, 109; blue, 109 }  ,draw opacity=1 ]   (148.27,64.6) .. controls (42.1,74.07) and (110.76,217.66) .. (185.05,156.56) ;
\draw [shift={(186.17,155.62)}, rotate = 499.65] [fill={rgb, 255:red, 109; green, 109; blue, 109 }  ,fill opacity=1 ][line width=0.75]  [draw opacity=0] (10.72,-5.15) -- (0,0) -- (10.72,5.15) -- (7.12,0) -- cycle    ;

\draw [color={rgb, 255:red, 109; green, 109; blue, 109 }  ,draw opacity=1 ]   (148.27,64.6) .. controls (39.48,100.85) and (156.77,211.55) .. (185.51,102.88) ;
\draw [shift={(185.94,101.22)}, rotate = 464.09] [fill={rgb, 255:red, 109; green, 109; blue, 109 }  ,fill opacity=1 ][line width=0.75]  [draw opacity=0] (10.72,-5.15) -- (0,0) -- (10.72,5.15) -- (7.12,0) -- cycle    ;

\draw (203.5,95) node  [align=left] {100};
\draw (153.5,41) node  [align=left] {000};
\draw  [draw opacity=0][fill={rgb, 255:red, 255; green, 255; blue, 255 }  ,fill opacity=0.5 ]  (104, 45) circle [x radius= 19.8, y radius= 12.73]   ;
\draw (104,45) node  [align=left] {001};
\draw  [draw opacity=0][fill={rgb, 255:red, 255; green, 255; blue, 255 }  ,fill opacity=0.5 ]  (50.5, 74) circle [x radius= 21.21, y radius= 12.73]   ;
\draw (50.5,74) node  [align=left] {011};
\draw (35,167) node  [align=left] {010};
\draw (79,208) node  [align=left] {110};
\draw (162.5,206) node  [align=left] {111};
\draw (203.5,164) node  [align=left] {101};
\draw (237,145.42) node [scale=0.9]  {$Re\left\{e^{j\psi ( t)}\right\}$};
\draw (170.54,14) node [scale=0.9]  {$Im\left\{e^{j\psi ( t)}\right\}$};

\end{tikzpicture}}
	}
	\end{tabular}
\captionsetup{font=footnotesize}
	\caption{BER for bit sub-channels with the Gray mapping, $M=3$}
	\label{fig:gray_mapping}
\end{figure}
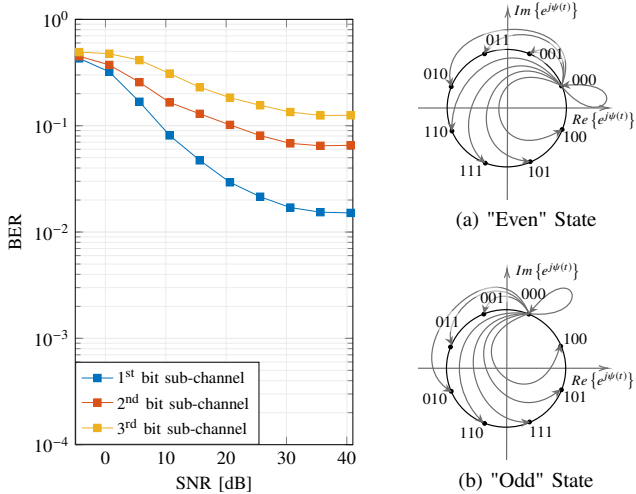
\subsection{Bit Mapping}
\begin{figure}[t!]
	\centering
	\begin{tabular}{ c c }
	\multirow{2}{*}[6.5em]{
	    \scalebox{.7}{
%
%
\definecolor{mycolor1}{rgb}{0.00000,0.44700,0.74100}%
\definecolor{mycolor2}{rgb}{0.85000,0.32500,0.09800}%
\definecolor{mycolor3}{rgb}{0.92900,0.69400,0.12500}%
\begin{tikzpicture}

\begin{axis}[%
width=15em,
height=23em,
scale only axis,
grid style={color=gray!15},
xmin=-5,
xmax=41,
xlabel={SNR [dB]},
ymode=log,
ymin=0.0001,
ymax=1,
yminorticks=true,
ylabel={BER},
xmajorgrids,
ymajorgrids,
yminorgrids,
legend style={at={(0,0)},anchor=south west,draw=black,fill=white,legend cell align=left,font=\small}
]
\addplot [smooth, color=mycolor1, mark=square*, mark options={solid, mycolor1}]
  table[row sep=crcr]{%
-4.3702302362906	0.44069696969697\\
0.629769763709402	0.349183333333333\\
5.6297697637094	0.215938069216758\\
10.6297697637094	0.105335335335335\\
15.6297697637094	0.0382231404958678\\
20.6297697637094	0.0121671216257204\\
25.6297697637094	0.00481546426185705\\
30.6297697637094	0.00255883434454863\\
35.6297697637094	0.00128870673952641\\
40.6297697637094	0.000354944718300053\\
};
\addlegendentry{$\text{1}^{\text{st}}\text{ bit sub-channel}$}

\addplot [smooth, color=mycolor2, mark=square*, mark options={solid, mycolor2}]
  table[row sep=crcr]{%
-4.3702302362906	0.457030303030303\\
0.629769763709402	0.37735\\
5.6297697637094	0.256803278688525\\
10.6297697637094	0.133353353353353\\
15.6297697637094	0.0518760330578512\\
20.6297697637094	0.016627236882014\\
25.6297697637094	0.00714929859719439\\
30.6297697637094	0.00357556536127965\\
35.6297697637094	0.00162644201578628\\
40.6297697637094	0.000380984209207118\\
};
\addlegendentry{$\text{2}^{\text{nd}}\text{ bit sub-channel}$}

\addplot [smooth, color=mycolor3, mark=square*, mark options={solid, mycolor3}]
  table[row sep=crcr]{%
-4.3702302362906	0.49130303030303\\
0.629769763709402	0.460183333333333\\
5.6297697637094	0.38051912568306\\
10.6297697637094	0.271566566566567\\
15.6297697637094	0.215564738292011\\
20.6297697637094	0.179774037003336\\
25.6297697637094	0.152442384769539\\
30.6297697637094	0.133159588159588\\
35.6297697637094	0.130415401740538\\
40.6297697637094	0.137342718689949\\
};
\addlegendentry{$\text{3}^{\text{rd}}\text{ bit sub-channel}$}

\end{axis}
\end{tikzpicture}%
}
	}
	&
    \subfloat["Even" State]{
        \scalebox{.6}{\tikzset{every picture/.style={line width=0.75pt}} 

\begin{tikzpicture}[x=0.75pt,y=0.75pt,yscale=-0.7,xscale=0.7]
\clip (66,0) rectangle + (260,240);

\draw [color={rgb, 255:red, 109; green, 109; blue, 109 }  ,draw opacity=1 ]   (173.13,250) -- (173.69,12.76) ;
\draw [shift={(173.7,10.76)}, rotate = 450.14] [color={rgb, 255:red, 109; green, 109; blue, 109 }  ,draw opacity=1 ][line width=0.75]    (10.93,-3.29) .. controls (6.95,-1.4) and (3.31,-0.3) .. (0,0) .. controls (3.31,0.3) and (6.95,1.4) .. (10.93,3.29)   ;

\draw [color={rgb, 255:red, 109; green, 109; blue, 109 }  ,draw opacity=1 ]   (52,130.38) -- (292.83,130.38) ;
\draw [shift={(294.83,130.38)}, rotate = 180] [color={rgb, 255:red, 109; green, 109; blue, 109 }  ,draw opacity=1 ][line width=0.75]    (10.93,-3.29) .. controls (6.95,-1.4) and (3.31,-0.3) .. (0,0) .. controls (3.31,0.3) and (6.95,1.4) .. (10.93,3.29)   ;

\draw   (100.96,131.09) .. controls (100.67,92.11) and (132.49,60.28) .. (172.02,60) .. controls (211.56,59.72) and (243.84,91.11) .. (244.13,130.09) .. controls (244.41,169.08) and (212.6,200.91) .. (173.06,201.19) .. controls (133.52,201.46) and (101.24,170.08) .. (100.96,131.09) -- cycle ;
\draw  [fill={rgb, 255:red, 0; green, 0; blue, 0 }  ,fill opacity=1 ] (143.64,65.31) .. controls (143.64,64.1) and (144.64,63.12) .. (145.86,63.12) .. controls (147.09,63.12) and (148.08,64.1) .. (148.08,65.31) .. controls (148.08,66.52) and (147.09,67.5) .. (145.86,67.5) .. controls (144.64,67.5) and (143.64,66.52) .. (143.64,65.31) -- cycle ;
\draw  [fill={rgb, 255:red, 0; green, 0; blue, 0 }  ,fill opacity=1 ] (197.61,65.31) .. controls (197.61,64.1) and (198.61,63.12) .. (199.83,63.12) .. controls (201.06,63.12) and (202.05,64.1) .. (202.05,65.31) .. controls (202.05,66.52) and (201.06,67.5) .. (199.83,67.5) .. controls (198.61,67.5) and (197.61,66.52) .. (197.61,65.31) -- cycle ;
\draw  [fill={rgb, 255:red, 0; green, 0; blue, 0 }  ,fill opacity=1 ] (103.6,104.8) .. controls (103.6,103.59) and (104.59,102.61) .. (105.82,102.61) .. controls (107.04,102.61) and (108.04,103.59) .. (108.04,104.8) .. controls (108.04,106.01) and (107.04,106.99) .. (105.82,106.99) .. controls (104.59,106.99) and (103.6,106.01) .. (103.6,104.8) -- cycle ;
\draw  [fill={rgb, 255:red, 0; green, 0; blue, 0 }  ,fill opacity=1 ] (104.47,158.02) .. controls (104.47,156.81) and (105.46,155.83) .. (106.69,155.83) .. controls (107.91,155.83) and (108.91,156.81) .. (108.91,158.02) .. controls (108.91,159.23) and (107.91,160.21) .. (106.69,160.21) .. controls (105.46,160.21) and (104.47,159.23) .. (104.47,158.02) -- cycle ;
\draw  [fill={rgb, 255:red, 0; green, 0; blue, 0 }  ,fill opacity=1 ] (235.91,103.94) .. controls (235.91,102.73) and (236.91,101.75) .. (238.13,101.75) .. controls (239.36,101.75) and (240.35,102.73) .. (240.35,103.94) .. controls (240.35,105.15) and (239.36,106.13) .. (238.13,106.13) .. controls (236.91,106.13) and (235.91,105.15) .. (235.91,103.94) -- cycle ;
\draw  [fill={rgb, 255:red, 0; green, 0; blue, 0 }  ,fill opacity=1 ] (144.51,196.65) .. controls (144.51,195.44) and (145.51,194.46) .. (146.73,194.46) .. controls (147.96,194.46) and (148.95,195.44) .. (148.95,196.65) .. controls (148.95,197.86) and (147.96,198.84) .. (146.73,198.84) .. controls (145.51,198.84) and (144.51,197.86) .. (144.51,196.65) -- cycle ;
\draw  [fill={rgb, 255:red, 0; green, 0; blue, 0 }  ,fill opacity=1 ] (198.48,194.93) .. controls (198.48,193.72) and (199.48,192.74) .. (200.7,192.74) .. controls (201.93,192.74) and (202.92,193.72) .. (202.92,194.93) .. controls (202.92,196.14) and (201.93,197.12) .. (200.7,197.12) .. controls (199.48,197.12) and (198.48,196.14) .. (198.48,194.93) -- cycle ;
\draw  [fill={rgb, 255:red, 0; green, 0; blue, 0 }  ,fill opacity=1 ] (236.78,156.3) .. controls (236.78,155.09) and (237.78,154.11) .. (239,154.11) .. controls (240.23,154.11) and (241.22,155.09) .. (241.22,156.3) .. controls (241.22,157.51) and (240.23,158.49) .. (239,158.49) .. controls (237.78,158.49) and (236.78,157.51) .. (236.78,156.3) -- cycle ;
\draw [color={rgb, 255:red, 109; green, 109; blue, 109 }  ,draw opacity=1 ]   (238.13,103.94) .. controls (289.02,166.93) and (328.92,92.46) .. (239.62,103.6) ;
\draw [shift={(238.27,103.77)}, rotate = 352.45] [fill={rgb, 255:red, 109; green, 109; blue, 109 }  ,fill opacity=1 ][line width=0.75]  [draw opacity=0] (10.72,-5.15) -- (0,0) -- (10.72,5.15) -- (7.12,0) -- cycle    ;

\draw [color={rgb, 255:red, 109; green, 109; blue, 109 }  ,draw opacity=1 ]   (238.13,103.94) .. controls (247.3,75.97) and (225.75,50.67) .. (201.33,64.42) ;
\draw [shift={(199.83,65.31)}, rotate = 327.95] [fill={rgb, 255:red, 109; green, 109; blue, 109 }  ,fill opacity=1 ][line width=0.75]  [draw opacity=0] (10.72,-5.15) -- (0,0) -- (10.72,5.15) -- (7.12,0) -- cycle    ;

\draw [color={rgb, 255:red, 109; green, 109; blue, 109 }  ,draw opacity=1 ]   (238.27,103.77) .. controls (258.5,36.08) and (165.06,26.74) .. (146.65,63.59) ;
\draw [shift={(145.86,65.31)}, rotate = 292.79] [fill={rgb, 255:red, 109; green, 109; blue, 109 }  ,fill opacity=1 ][line width=0.75]  [draw opacity=0] (10.72,-5.15) -- (0,0) -- (10.72,5.15) -- (7.12,0) -- cycle    ;

\draw [color={rgb, 255:red, 109; green, 109; blue, 109 }  ,draw opacity=1 ]   (238.13,103.94) .. controls (276.89,9.77) and (85.77,16.95) .. (105.5,103.49) ;
\draw [shift={(105.82,104.8)}, rotate = 255.94] [fill={rgb, 255:red, 109; green, 109; blue, 109 }  ,fill opacity=1 ][line width=0.75]  [draw opacity=0] (10.72,-5.15) -- (0,0) -- (10.72,5.15) -- (7.12,0) -- cycle    ;

\draw [color={rgb, 255:red, 109; green, 109; blue, 109 }  ,draw opacity=1 ]   (238.13,103.94) .. controls (182.48,31.98) and (104.64,87.71) .. (106.65,156.97) ;
\draw [shift={(106.69,158.02)}, rotate = 267.68] [fill={rgb, 255:red, 109; green, 109; blue, 109 }  ,fill opacity=1 ][line width=0.75]  [draw opacity=0] (10.72,-5.15) -- (0,0) -- (10.72,5.15) -- (7.12,0) -- cycle    ;

\draw [color={rgb, 255:red, 109; green, 109; blue, 109 }  ,draw opacity=1 ]   (238.13,103.94) .. controls (166.89,29.42) and (89.84,146.32) .. (145.88,195.91) ;
\draw [shift={(146.73,196.65)}, rotate = 220.44] [fill={rgb, 255:red, 109; green, 109; blue, 109 }  ,fill opacity=1 ][line width=0.75]  [draw opacity=0] (10.72,-5.15) -- (0,0) -- (10.72,5.15) -- (7.12,0) -- cycle    ;

\draw [color={rgb, 255:red, 109; green, 109; blue, 109 }  ,draw opacity=1 ]   (238.27,103.77) .. controls (156.5,35.39) and (103.52,185.48) .. (199.25,194.8) ;
\draw [shift={(200.7,194.93)}, rotate = 184.65] [fill={rgb, 255:red, 109; green, 109; blue, 109 }  ,fill opacity=1 ][line width=0.75]  [draw opacity=0] (10.72,-5.15) -- (0,0) -- (10.72,5.15) -- (7.12,0) -- cycle    ;

\draw [color={rgb, 255:red, 109; green, 109; blue, 109 }  ,draw opacity=1 ]   (238.27,103.77) .. controls (135.71,52.48) and (140.37,213.69) .. (237.53,157.17) ;
\draw [shift={(239,156.3)}, rotate = 509.09] [fill={rgb, 255:red, 109; green, 109; blue, 109 }  ,fill opacity=1 ][line width=0.75]  [draw opacity=0] (10.72,-5.15) -- (0,0) -- (10.72,5.15) -- (7.12,0) -- cycle    ;

\draw (265.51,96.69) node  [align=left] {000};
\draw  [draw opacity=0][fill={rgb, 255:red, 255; green, 255; blue, 255 }  ,fill opacity=0.5 ]  (226.34, 64.92) circle [x radius= 21.21, y radius= 12.73]   ;
\draw (226.34,64.92) node  [align=left] {001};
\draw  [draw opacity=0][fill={rgb, 255:red, 255; green, 255; blue, 255 }  ,fill opacity=0.5 ]  (161.92, 48.61) circle [x radius= 19.8, y radius= 12.73]   ;
\draw (161.92,48.61) node  [align=left] {011};
\draw (87.06,88.96) node  [align=left] {010};
\draw (87.06,159.35) node  [align=left] {110};
\draw (129.71,205.71) node  [align=left] {111};
\draw (211.54,207.42) node  [align=left] {101};
\draw (254.19,164.5) node  [align=left] {100};
\draw (289,145.42) node [scale=0.9]  {$Re\left\{e^{j\psi ( t)}\right\}$};
\draw (220,14) node [scale=0.9]  {$Im\left\{e^{j\psi ( t)}\right\}$};

\end{tikzpicture}}
    }

	\\
	&
	\subfloat["Odd" State]{
	    \scalebox{.6}{\tikzset{every picture/.style={line width=0.75pt}} 

\begin{tikzpicture}[x=0.75pt,y=0.75pt,yscale=-0.7,xscale=0.7]
\clip (15,0) rectangle + (260,240);

\draw [color={rgb, 255:red, 109; green, 109; blue, 109 }  ,draw opacity=1 ]   (121.13,250) -- (121.69,12.76) ;
\draw [shift={(121.7,10.76)}, rotate = 450.14] [color={rgb, 255:red, 109; green, 109; blue, 109 }  ,draw opacity=1 ][line width=0.75]    (10.93,-3.29) .. controls (6.95,-1.4) and (3.31,-0.3) .. (0,0) .. controls (3.31,0.3) and (6.95,1.4) .. (10.93,3.29)   ;

\draw [color={rgb, 255:red, 109; green, 109; blue, 109 }  ,draw opacity=1 ]   (0,130.38) -- (240.83,130.38) ;
\draw [shift={(242.83,130.38)}, rotate = 180] [color={rgb, 255:red, 109; green, 109; blue, 109 }  ,draw opacity=1 ][line width=0.75]    (10.93,-3.29) .. controls (6.95,-1.4) and (3.31,-0.3) .. (0,0) .. controls (3.31,0.3) and (6.95,1.4) .. (10.93,3.29)   ;

\draw   (48.96,131.09) .. controls (48.67,92.11) and (80.49,60.28) .. (120.02,60) .. controls (159.56,59.72) and (191.84,91.11) .. (192.13,130.09) .. controls (192.41,169.08) and (160.6,200.91) .. (121.06,201.19) .. controls (81.52,201.46) and (49.24,170.08) .. (48.96,131.09) -- cycle ;
\draw  [fill={rgb, 255:red, 0; green, 0; blue, 0 }  ,fill opacity=1 ] (91.64,65.31) .. controls (91.64,64.1) and (92.64,63.12) .. (93.86,63.12) .. controls (95.09,63.12) and (96.08,64.1) .. (96.08,65.31) .. controls (96.08,66.52) and (95.09,67.5) .. (93.86,67.5) .. controls (92.64,67.5) and (91.64,66.52) .. (91.64,65.31) -- cycle ;
\draw  [fill={rgb, 255:red, 0; green, 0; blue, 0 }  ,fill opacity=1 ] (145.61,65.31) .. controls (145.61,64.1) and (146.61,63.12) .. (147.83,63.12) .. controls (149.06,63.12) and (150.05,64.1) .. (150.05,65.31) .. controls (150.05,66.52) and (149.06,67.5) .. (147.83,67.5) .. controls (146.61,67.5) and (145.61,66.52) .. (145.61,65.31) -- cycle ;
\draw  [fill={rgb, 255:red, 0; green, 0; blue, 0 }  ,fill opacity=1 ] (51.6,104.8) .. controls (51.6,103.59) and (52.59,102.61) .. (53.82,102.61) .. controls (55.04,102.61) and (56.04,103.59) .. (56.04,104.8) .. controls (56.04,106.01) and (55.04,106.99) .. (53.82,106.99) .. controls (52.59,106.99) and (51.6,106.01) .. (51.6,104.8) -- cycle ;
\draw  [fill={rgb, 255:red, 0; green, 0; blue, 0 }  ,fill opacity=1 ] (52.47,158.02) .. controls (52.47,156.81) and (53.46,155.83) .. (54.69,155.83) .. controls (55.91,155.83) and (56.91,156.81) .. (56.91,158.02) .. controls (56.91,159.23) and (55.91,160.21) .. (54.69,160.21) .. controls (53.46,160.21) and (52.47,159.23) .. (52.47,158.02) -- cycle ;
\draw  [fill={rgb, 255:red, 0; green, 0; blue, 0 }  ,fill opacity=1 ] (183.91,103.94) .. controls (183.91,102.73) and (184.91,101.75) .. (186.13,101.75) .. controls (187.36,101.75) and (188.35,102.73) .. (188.35,103.94) .. controls (188.35,105.15) and (187.36,106.13) .. (186.13,106.13) .. controls (184.91,106.13) and (183.91,105.15) .. (183.91,103.94) -- cycle ;
\draw  [fill={rgb, 255:red, 0; green, 0; blue, 0 }  ,fill opacity=1 ] (92.51,196.65) .. controls (92.51,195.44) and (93.51,194.46) .. (94.73,194.46) .. controls (95.96,194.46) and (96.95,195.44) .. (96.95,196.65) .. controls (96.95,197.86) and (95.96,198.84) .. (94.73,198.84) .. controls (93.51,198.84) and (92.51,197.86) .. (92.51,196.65) -- cycle ;
\draw  [fill={rgb, 255:red, 0; green, 0; blue, 0 }  ,fill opacity=1 ] (146.48,194.93) .. controls (146.48,193.72) and (147.48,192.74) .. (148.7,192.74) .. controls (149.93,192.74) and (150.92,193.72) .. (150.92,194.93) .. controls (150.92,196.14) and (149.93,197.12) .. (148.7,197.12) .. controls (147.48,197.12) and (146.48,196.14) .. (146.48,194.93) -- cycle ;
\draw  [fill={rgb, 255:red, 0; green, 0; blue, 0 }  ,fill opacity=1 ] (184.78,156.3) .. controls (184.78,155.09) and (185.78,154.11) .. (187,154.11) .. controls (188.23,154.11) and (189.22,155.09) .. (189.22,156.3) .. controls (189.22,157.51) and (188.23,158.49) .. (187,158.49) .. controls (185.78,158.49) and (184.78,157.51) .. (184.78,156.3) -- cycle ;
\draw [color={rgb, 255:red, 109; green, 109; blue, 109 }  ,draw opacity=1 ]   (148.29,64.81) .. controls (228.82,73.37) and (204.37,-7.5) .. (149.11,63.52) ;
\draw [shift={(148.27,64.6)}, rotate = 307.45] [fill={rgb, 255:red, 109; green, 109; blue, 109 }  ,fill opacity=1 ][line width=0.75]  [draw opacity=0] (10.72,-5.15) -- (0,0) -- (10.72,5.15) -- (7.12,0) -- cycle    ;

\draw [color={rgb, 255:red, 109; green, 109; blue, 109 }  ,draw opacity=1 ]   (148.29,64.81) .. controls (135,38.55) and (101.87,35.91) .. (94.32,62.89) ;
\draw [shift={(93.9,64.58)}, rotate = 282.95] [fill={rgb, 255:red, 109; green, 109; blue, 109 }  ,fill opacity=1 ][line width=0.75]  [draw opacity=0] (10.72,-5.15) -- (0,0) -- (10.72,5.15) -- (7.12,0) -- cycle    ;

\draw [color={rgb, 255:red, 109; green, 109; blue, 109 }  ,draw opacity=1 ]   (148.27,64.6) .. controls (114.71,2.43) and (42.03,61.9) .. (55.08,100.97) ;
\draw [shift={(55.73,102.75)}, rotate = 247.79000000000002] [fill={rgb, 255:red, 109; green, 109; blue, 109 }  ,fill opacity=1 ][line width=0.75]  [draw opacity=0] (10.72,-5.15) -- (0,0) -- (10.72,5.15) -- (7.12,0) -- cycle    ;

\draw [color={rgb, 255:red, 109; green, 109; blue, 109 }  ,draw opacity=1 ]   (148.29,64.81) .. controls (109.11,-29.18) and (-20.96,111.04) .. (54.19,158.28) ;
\draw [shift={(55.34,158.98)}, rotate = 210.94] [fill={rgb, 255:red, 109; green, 109; blue, 109 }  ,fill opacity=1 ][line width=0.75]  [draw opacity=0] (10.72,-5.15) -- (0,0) -- (10.72,5.15) -- (7.12,0) -- cycle    ;

\draw [color={rgb, 255:red, 109; green, 109; blue, 109 }  ,draw opacity=1 ]   (148.29,64.81) .. controls (58.06,53.28) and (42.42,147.73) .. (92.82,195.29) ;
\draw [shift={(93.59,196)}, rotate = 222.68] [fill={rgb, 255:red, 109; green, 109; blue, 109 }  ,fill opacity=1 ][line width=0.75]  [draw opacity=0] (10.72,-5.15) -- (0,0) -- (10.72,5.15) -- (7.12,0) -- cycle    ;

\draw [color={rgb, 255:red, 109; green, 109; blue, 109 }  ,draw opacity=1 ]   (148.29,64.81) .. controls (45.22,62.49) and (73.4,199.65) .. (148.09,195.08) ;
\draw [shift={(149.22,195)}, rotate = 535.44] [fill={rgb, 255:red, 109; green, 109; blue, 109 }  ,fill opacity=1 ][line width=0.75]  [draw opacity=0] (10.72,-5.15) -- (0,0) -- (10.72,5.15) -- (7.12,0) -- cycle    ;

\draw [color={rgb, 255:red, 109; green, 109; blue, 109 }  ,draw opacity=1 ]   (148.27,64.6) .. controls (42.1,74.07) and (110.76,217.66) .. (185.05,156.56) ;
\draw [shift={(186.17,155.62)}, rotate = 499.65] [fill={rgb, 255:red, 109; green, 109; blue, 109 }  ,fill opacity=1 ][line width=0.75]  [draw opacity=0] (10.72,-5.15) -- (0,0) -- (10.72,5.15) -- (7.12,0) -- cycle    ;

\draw [color={rgb, 255:red, 109; green, 109; blue, 109 }  ,draw opacity=1 ]   (148.27,64.6) .. controls (39.48,100.85) and (156.77,211.55) .. (185.51,102.88) ;
\draw [shift={(185.94,101.22)}, rotate = 464.09] [fill={rgb, 255:red, 109; green, 109; blue, 109 }  ,fill opacity=1 ][line width=0.75]  [draw opacity=0] (10.72,-5.15) -- (0,0) -- (10.72,5.15) -- (7.12,0) -- cycle    ;

\draw (203.5,95) node  [align=left] {000};
\draw (153.5,41) node  [align=left] {001};
\draw  [draw opacity=0][fill={rgb, 255:red, 255; green, 255; blue, 255 }  ,fill opacity=0.5 ]  (104, 45) circle [x radius= 19.8, y radius= 12.73]   ;
\draw (104,45) node  [align=left] {011};
\draw  [draw opacity=0][fill={rgb, 255:red, 255; green, 255; blue, 255 }  ,fill opacity=0.5 ]  (50.5, 74) circle [x radius= 21.21, y radius= 12.73]   ;
\draw (50.5,74) node  [align=left] {010};
\draw (35,167) node  [align=left] {110};
\draw (79,208) node  [align=left] {111};
\draw (162.5,206) node  [align=left] {101};
\draw (203.5,164) node  [align=left] {100};
\draw (237,145.42) node [scale=0.9]  {$Re\left\{e^{j\psi ( t)}\right\}$};
\draw (170,14) node [scale=0.9]  {$Im\left\{e^{j\psi ( t)}\right\}$};

\end{tikzpicture}}
	}
	\end{tabular}
\captionsetup{font=footnotesize}
\caption{BER for bit sub-channels with the Advanced mapping, $M=3$}
	\label{fig:advanced_mapping}
\end{figure}
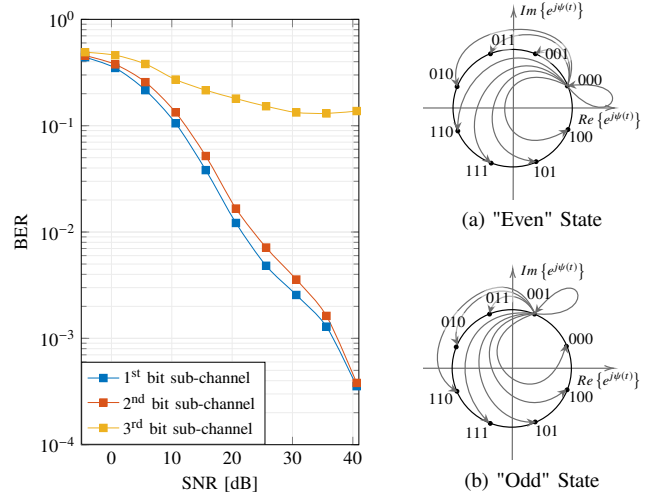

The established Gray mapping scheme implies well known benefits for conventional CPM systems.
However, in the sequel it is proposed to  modify the Gray coding scheme in order to enable the exploitation of the properties of the CPM system with 1-bit quantization and oversampling at the receiver. 
The novel mapping scheme, termed advanced mapping, allows for the separation of the information that can be readily extracted from the orthants of the symbols and the additional information which is brought by higher-order modulation and the timing of the transitions between the orthants. 
In this regard, this study proposes a novel state depending, bit mapping, which allows to divide the $\log_2(M_{\textrm{cpm}})=3$ bits into binary sub-channels where two sub-channels can each yield up to 1 bit per channel use and the third sub-channel yields a lower achievable rate which depends on the oversampling factor.
The considered mappings are presented in Fig.~\ref{fig:gray_mapping} and Fig.~\ref{fig:advanced_mapping}, which show BER performance of each bit sub-channel, as well as the bit allocations for the possible state transitions, given an "even" or an "odd" initial state, whose parity is defined by the parity of the absolute phase state described in \eqref{eq:cpm:beta}. Note that the advanced mapping corresponds to a circular shift into the Gray mapping when the initial state of the transition is "odd". With this, the uncertainties brought by the coarse quantization are concentrated into the third bit sub-channel.

\subsection{Subchannel Coding Scheme}
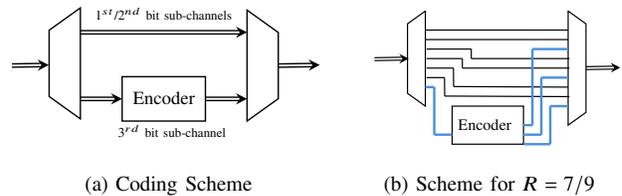
\begin{figure}[t]
{	\centering{
	\subfloat[Coding Scheme]{
	    \scalebox{.7}{\tikzset{every picture/.style={line width=0.75pt}} 

\begin{tikzpicture}[x=0.75pt,y=0.75pt,yscale=-0.8,xscale=0.75]

\clip (5,10) rectangle + (300,150);

\draw   (70,120) -- (40,99.84) -- (40,40.16) -- (70,20) -- cycle ;
\draw    (1.11,70.5) -- (31.64,70.04)(1.16,73.5) -- (31.69,73.04) ;
\draw [shift={(39.67,71.42)}, rotate = 539.13] [fill={rgb, 255:red, 0; green, 0; blue, 0 }  ][line width=0.75]  [draw opacity=0] (10.72,-5.15) -- (0,0) -- (10.72,5.15) -- (7.12,0) -- cycle    ;

\draw    (70.13,40.5) -- (222.13,40.5)(70.13,43.5) -- (222.13,43.5) ;
\draw [shift={(230.13,42)}, rotate = 180] [fill={rgb, 255:red, 0; green, 0; blue, 0 }  ][line width=0.75]  [draw opacity=0] (10.72,-5.15) -- (0,0) -- (10.72,5.15) -- (7.12,0) -- cycle    ;

\draw    (70.13,100.5) -- (102.13,100.5)(70.13,103.5) -- (102.13,103.5) ;
\draw [shift={(110.13,102)}, rotate = 180] [fill={rgb, 255:red, 0; green, 0; blue, 0 }  ][line width=0.75]  [draw opacity=0] (10.72,-5.15) -- (0,0) -- (10.72,5.15) -- (7.12,0) -- cycle    ;

\draw   (110.13,82) -- (190.13,82) -- (190.13,122) -- (110.13,122) -- cycle ;

\draw   (230.13,22) -- (260.13,42.16) -- (260.13,101.84) -- (230.13,122) -- cycle ;
\draw    (190.13,100.5) -- (222.13,100.5)(190.13,103.5) -- (222.13,103.5) ;
\draw [shift={(230.13,102)}, rotate = 180] [fill={rgb, 255:red, 0; green, 0; blue, 0 }  ][line width=0.75]  [draw opacity=0] (10.72,-5.15) -- (0,0) -- (10.72,5.15) -- (7.12,0) -- cycle    ;

\draw    (260.11,70.5) -- (291.98,70.03)(260.16,73.5) -- (292.02,73.03) ;
\draw [shift={(300,71.42)}, rotate = 539.1600000000001] [fill={rgb, 255:red, 0; green, 0; blue, 0 }  ][line width=0.75]  [draw opacity=0] (10.72,-5.15) -- (0,0) -- (10.72,5.15) -- (7.12,0) -- cycle    ;

\draw   (378.53,110) -- (358.53,96.56) -- (358.53,23.44) -- (378.53,10) -- cycle ;
\draw    (321.44,59.08) -- (351.98,58.62)(321.49,62.08) -- (352.02,61.62) ;
\draw [shift={(360,60)}, rotate = 539.13] [fill={rgb, 255:red, 0; green, 0; blue, 0 }  ][line width=0.75]  [draw opacity=0] (10.72,-5.15) -- (0,0) -- (10.72,5.15) -- (7.12,0) -- cycle    ;

\draw   (538.53,10) -- (558.53,23.44) -- (558.53,116.56) -- (538.53,130) -- cycle ;
\draw   (408.53,110) -- (488.53,110) -- (488.53,150) -- (408.53,150) -- cycle ;

\draw    (538.53,30) -- (378.53,30) ;

\draw    (538.53,40) -- (378.53,40) ;

\draw    (540,60) -- (430,60) ;

\draw    (540,70) -- (420,70) ;

\draw    (538.53,90) -- (410.2,89.42) ;

\draw    (540,100) -- (400.2,99.42) ;

\draw [color={rgb, 255:red, 74; green, 144; blue, 226 }  ,draw opacity=1 ][line width=1.5]    (408.53,140) -- (388.53,140) ;

\draw [color={rgb, 255:red, 74; green, 144; blue, 226 }  ,draw opacity=1 ][line width=1.5]    (498.53,130) -- (488.53,130) ;

\draw [color={rgb, 255:red, 74; green, 144; blue, 226 }  ,draw opacity=1 ][line width=1.5]    (498.53,130) -- (498.53,50) ;

\draw [color={rgb, 255:red, 74; green, 144; blue, 226 }  ,draw opacity=1 ][line width=1.5]    (498.53,50) -- (538.53,50) ;

\draw [color={rgb, 255:red, 74; green, 144; blue, 226 }  ,draw opacity=1 ][line width=1.5]    (508.53,140) -- (488.53,140) ;

\draw [color={rgb, 255:red, 74; green, 144; blue, 226 }  ,draw opacity=1 ][line width=1.5]    (508.53,140) -- (508.53,80) ;

\draw [color={rgb, 255:red, 74; green, 144; blue, 226 }  ,draw opacity=1 ][line width=1.5]    (538.53,80) -- (508.53,80) ;

\draw [color={rgb, 255:red, 74; green, 144; blue, 226 }  ,draw opacity=1 ][line width=1.5]    (518.53,150) -- (488.53,150) ;

\draw [color={rgb, 255:red, 74; green, 144; blue, 226 }  ,draw opacity=1 ][line width=1.5]    (538.53,110) -- (518.53,110) ;

\draw [color={rgb, 255:red, 74; green, 144; blue, 226 }  ,draw opacity=1 ][line width=1.5]    (518.53,150) -- (518.53,110) ;

\draw    (558.51,68.5) -- (589.04,68.04)(558.56,71.5) -- (589.09,71.04) ;
\draw [shift={(597.07,69.42)}, rotate = 539.13] [fill={rgb, 255:red, 0; green, 0; blue, 0 }  ][line width=0.75]  [draw opacity=0] (10.72,-5.15) -- (0,0) -- (10.72,5.15) -- (7.12,0) -- cycle    ;

\draw    (420,70) -- (420,60) ;

\draw    (430,50) -- (380,50) ;

\draw    (430,60) -- (430,50) ;

\draw    (420,60) -- (380,60) ;

\draw    (380,80) -- (400,80) ;

\draw    (380,70) -- (410,70) ;

\draw    (410.2,89.42) -- (410,70) ;

\draw    (400.2,99.42) -- (400,80) ;

\draw [color={rgb, 255:red, 74; green, 144; blue, 226 }  ,draw opacity=1 ][line width=1.5]    (388.53,90) -- (378.53,90) ;

\draw [color={rgb, 255:red, 74; green, 144; blue, 226 }  ,draw opacity=1 ][line width=1.5]    (388.53,140) -- (388.53,90) ;

\draw (721,21) node   {$0$};
\draw (701,71) node   {$0$};
\draw (150.13,101) node  [align=left] {Encoder};
\draw (158.13,131) node  [align=left] {{\scriptsize $3^{rd}$ bit sub-channel}};
\draw (150,25) node  [align=left] {{\scriptsize $1^{st}$/$2^{nd}$ bit sub-channels}};
\draw (445,129) node  [align=left] {Encoder};

\end{tikzpicture}}
	}
	\subfloat[Scheme for $R=7/9$]{
	    \scalebox{.6}{\tikzset{every picture/.style={line width=0.75pt}} 

\begin{tikzpicture}[x=0.75pt,y=0.75pt,yscale=-0.8,xscale=0.75]

\clip (300,10) rectangle + (300,160);

\draw   (70,120) -- (40,99.84) -- (40,40.16) -- (70,20) -- cycle ;
\draw    (1.11,70.5) -- (31.64,70.04)(1.16,73.5) -- (31.69,73.04) ;
\draw [shift={(39.67,71.42)}, rotate = 539.13] [fill={rgb, 255:red, 0; green, 0; blue, 0 }  ][line width=0.75]  [draw opacity=0] (10.72,-5.15) -- (0,0) -- (10.72,5.15) -- (7.12,0) -- cycle    ;

\draw    (70.13,40.5) -- (222.13,40.5)(70.13,43.5) -- (222.13,43.5) ;
\draw [shift={(230.13,42)}, rotate = 180] [fill={rgb, 255:red, 0; green, 0; blue, 0 }  ][line width=0.75]  [draw opacity=0] (10.72,-5.15) -- (0,0) -- (10.72,5.15) -- (7.12,0) -- cycle    ;

\draw    (70.13,100.5) -- (102.13,100.5)(70.13,103.5) -- (102.13,103.5) ;
\draw [shift={(110.13,102)}, rotate = 180] [fill={rgb, 255:red, 0; green, 0; blue, 0 }  ][line width=0.75]  [draw opacity=0] (10.72,-5.15) -- (0,0) -- (10.72,5.15) -- (7.12,0) -- cycle    ;

\draw   (110.13,82) -- (190.13,82) -- (190.13,122) -- (110.13,122) -- cycle ;

\draw   (230.13,22) -- (260.13,42.16) -- (260.13,101.84) -- (230.13,122) -- cycle ;
\draw    (190.13,100.5) -- (222.13,100.5)(190.13,103.5) -- (222.13,103.5) ;
\draw [shift={(230.13,102)}, rotate = 180] [fill={rgb, 255:red, 0; green, 0; blue, 0 }  ][line width=0.75]  [draw opacity=0] (10.72,-5.15) -- (0,0) -- (10.72,5.15) -- (7.12,0) -- cycle    ;

\draw    (260.11,70.5) -- (291.98,70.03)(260.16,73.5) -- (292.02,73.03) ;
\draw [shift={(300,71.42)}, rotate = 539.1600000000001] [fill={rgb, 255:red, 0; green, 0; blue, 0 }  ][line width=0.75]  [draw opacity=0] (10.72,-5.15) -- (0,0) -- (10.72,5.15) -- (7.12,0) -- cycle    ;

\draw   (378.53,110) -- (358.53,96.56) -- (358.53,23.44) -- (378.53,10) -- cycle ;
\draw    (321.44,59.08) -- (351.98,58.62)(321.49,62.08) -- (352.02,61.62) ;
\draw [shift={(360,60)}, rotate = 539.13] [fill={rgb, 255:red, 0; green, 0; blue, 0 }  ][line width=0.75]  [draw opacity=0] (10.72,-5.15) -- (0,0) -- (10.72,5.15) -- (7.12,0) -- cycle    ;

\draw   (538.53,10) -- (558.53,23.44) -- (558.53,116.56) -- (538.53,130) -- cycle ;
\draw   (408.53,110) -- (488.53,110) -- (488.53,150) -- (408.53,150) -- cycle ;

\draw    (538.53,30) -- (378.53,30) ;

\draw    (538.53,40) -- (378.53,40) ;

\draw    (540,60) -- (430,60) ;

\draw    (540,70) -- (420,70) ;

\draw    (538.53,90) -- (410.2,89.42) ;

\draw    (540,100) -- (400.2,99.42) ;

\draw [color={rgb, 255:red, 74; green, 144; blue, 226 }  ,draw opacity=1 ][line width=1.5]    (408.53,140) -- (388.53,140) ;

\draw [color={rgb, 255:red, 74; green, 144; blue, 226 }  ,draw opacity=1 ][line width=1.5]    (498.53,130) -- (488.53,130) ;

\draw [color={rgb, 255:red, 74; green, 144; blue, 226 }  ,draw opacity=1 ][line width=1.5]    (498.53,130) -- (498.53,50) ;

\draw [color={rgb, 255:red, 74; green, 144; blue, 226 }  ,draw opacity=1 ][line width=1.5]    (498.53,50) -- (538.53,50) ;

\draw [color={rgb, 255:red, 74; green, 144; blue, 226 }  ,draw opacity=1 ][line width=1.5]    (508.53,140) -- (488.53,140) ;

\draw [color={rgb, 255:red, 74; green, 144; blue, 226 }  ,draw opacity=1 ][line width=1.5]    (508.53,140) -- (508.53,80) ;

\draw [color={rgb, 255:red, 74; green, 144; blue, 226 }  ,draw opacity=1 ][line width=1.5]    (538.53,80) -- (508.53,80) ;

\draw [color={rgb, 255:red, 74; green, 144; blue, 226 }  ,draw opacity=1 ][line width=1.5]    (518.53,150) -- (488.53,150) ;

\draw [color={rgb, 255:red, 74; green, 144; blue, 226 }  ,draw opacity=1 ][line width=1.5]    (538.53,110) -- (518.53,110) ;

\draw [color={rgb, 255:red, 74; green, 144; blue, 226 }  ,draw opacity=1 ][line width=1.5]    (518.53,150) -- (518.53,110) ;

\draw    (558.51,68.5) -- (589.04,68.04)(558.56,71.5) -- (589.09,71.04) ;
\draw [shift={(597.07,69.42)}, rotate = 539.13] [fill={rgb, 255:red, 0; green, 0; blue, 0 }  ][line width=0.75]  [draw opacity=0] (10.72,-5.15) -- (0,0) -- (10.72,5.15) -- (7.12,0) -- cycle    ;

\draw    (420,70) -- (420,60) ;

\draw    (430,50) -- (380,50) ;

\draw    (430,60) -- (430,50) ;

\draw    (420,60) -- (380,60) ;

\draw    (380,80) -- (400,80) ;

\draw    (380,70) -- (410,70) ;

\draw    (410.2,89.42) -- (410,70) ;

\draw    (400.2,99.42) -- (400,80) ;

\draw [color={rgb, 255:red, 74; green, 144; blue, 226 }  ,draw opacity=1 ][line width=1.5]    (388.53,90) -- (378.53,90) ;

\draw [color={rgb, 255:red, 74; green, 144; blue, 226 }  ,draw opacity=1 ][line width=1.5]    (388.53,140) -- (388.53,90) ;

\draw (721,21) node   {$0$};
\draw (701,71) node   {$0$};
\draw (150.13,101) node  [align=left] {Encoder};
\draw (158.13,131) node  [align=left] {{\scriptsize $3^{rd}$ bit sub-channel}};
\draw (150,25) node  [align=left] {{\scriptsize $1^{st}$ / $2^{nd}$ bit sub-channels}};
\draw (445,129) node  [align=left] {Encoder};

\end{tikzpicture}}
	}}}
\captionsetup{font=footnotesize}
\caption{Proposed Coding Scheme for $M_{\textrm{cpm}}=8$}
	\label{fig:prop_coding}
\end{figure}
The coding scheme which we refer to as the conventional approach, consists of the system model illustrated in Fig.~\ref{fig:discrete_system}. Such a scheme uses convolutional codes as forward error correction with no regard towards bit sub-channel performance. More over Interleavers are used to protect the system against burst errors and an iterative decoding procedure is applied. On the other hand, the proposed coding scheme has the same structure, but takes into account the Advanced mapping to apply different code rates for the bit sub-channels. For the case study, no coding is applied to the first two bit sub-channels whereas a strong convolutional code is applied to the third sub-channel. This is illustrated in Fig.~\ref{fig:prop_coding}, where in (b) a code rate of $1/3$ is applied to the third sub-channel, which corresponds to an overall code rate of $R=7/9$.

\section{Numerical Results}
\label{sec:numerical_results}
All the computations rely on $M_{\textrm{cpm}}=8$ and modulation index $h=\frac{1}{M_{\textrm{cpm}}}$. The used frequency pulse is the 1REC \cite{Anderson_1986}.
A suboptimal bandpass filter is considered with
\begin{align}
g_{\textrm{IF}}(t)  =    \sqrt{ \frac{1}{T_g}}   \mathrm{rect}\left(\frac{t- T_{\textrm{s}} / 2  }{T_g}\right)  \cdot e^{j 2 \pi \Delta f (t-T_{\textrm{s}} / 2) }   \textrm{,}
\end{align}
where $T_g =  \frac{1}{2} T_{\textrm{s}}$. The extraction of the soft information is done based on an auxiliary channel law $W(\cdot)$ with the parameter $N=0$, cf.\ \eqref{eq:aux_channel}.
The $\mathrm{SNR}$ is expressed by the ratio between the transmit power and the noise power in the signal band, which reads as
$\mathrm{SNR}  =
E_\textrm{s} /    (  T_{\textrm{s}}
B_{90\%}  N_0 )$, where $B_{90\%}$ denotes the $90\%$ power containment bandwidth.


A block of information bits is randomly generated and forwarded to the encoder according to the discrete system model described in Fig.~\ref{fig:discrete_system}. In order to modify the code rate, puncturing is performed using the puncturing patterns described in Table \ref{table:punctcodes1}. S-Random interleavers are used such that they are generated for each analyzed block.

{\footnotesize
\begin{table}[h]
\centering
\scriptsize
	\begin{tabular}{ c c c }
	\hline
	Code Rate & Generator & Puncturing Pattern \\
	\hline
	$1/3$ & (5 7 7) & 1 1 1 \\
	$1/2$ & (5 7) & 1 1 \\
	$2/3$ & (5 7) & 1 1 $\vert$ 0 1 \\
	$3/4$ & (5 7) & 1 1 $\vert$ 0 1 $\vert$ 1 0 \\
	\hline
	\end{tabular}
	\captionsetup{font=footnotesize}
	\caption{Puncturing patterns for codes with Constraint Length $K_{cc}=3$, generator polynomial in octal form}
	\label{table:punctcodes1}
\end{table}
}


\begin{figure}[t]
	\centering
	\definecolor{mycolor1}{rgb}{0.80000,1.00000,0.00000}%
\definecolor{mycolor2}{rgb}{0.00000,1.00000,0.40000}%
\definecolor{mycolor3}{rgb}{0.00000,0.40000,1.00000}%
\definecolor{mycolor4}{rgb}{0.80000,0.00000,1.00000}%
\begin{tikzpicture}

\begin{axis}[%
width=0.82\columnwidth,
height=0.4375\columnwidth,
scale only axis,
grid style={color=gray!15},
xmin=-5,
xmax=41,
xlabel={SNR [dB]},
ymode=log,
ymin=5e-07,
ymax=0.99,
yminorticks=true,
ylabel={BER},
xmajorgrids,
ymajorgrids,
yminorgrids,
legend entries={
    \text{Gray Mapping ($3^{\textrm{rd}}$ iter.)},
    \text{Advanced Mapping ($2^{\textrm{nd}}$ iter.)}
},
legend style={at={(0,0)},anchor=south west,draw=black,fill=white,legend cell align=left,font=\tiny}
]

\addlegendimage{dashed, mark=o, mark options={solid}}
\addlegendimage{mark=triangle*}

\addplot [smooth, color=red, mark=square*, mark options={solid, red}, forget plot]
  table[row sep=crcr]{%
-4.3702302362906	0.463010101010101\\
0.629769763709402	0.395572222222222\\
5.6297697637094	0.284420157862781\\
10.6297697637094	0.170085085085085\\
15.6297697637094	0.101887970615243\\
20.6297697637094	0.0695227985036902\\
25.6297697637094	0.0548023825428635\\
30.6297697637094	0.0464313292884721\\
35.6297697637094	0.0444435168319503\\
40.6297697637094	0.0460262158724854\\
};

\addplot [smooth, color=mycolor1, mark=triangle*, mark options={solid, mycolor1}, forget plot]
  table[row sep=crcr]{%
-4.3702302362906	0.500606060606061\\
0.629769763709402	0.462633333333333\\
5.6297697637094	0.291183970856102\\
10.6297697637094	0.0256056056056056\\
15.6297697637094	0.00105234159779614\\
20.6297697637094	0.00010919017288444\\
25.6297697637094	1.1690046760187e-05\\
30.6297697637094	6.89464975179261e-06\\
35.6297697637094	2.78283748229103e-06\\
40.6297697637094	8.35491686857716e-07\\
};

\addplot [smooth, color=mycolor2, mark=triangle*, mark options={solid, mycolor2}, forget plot]
  table[row sep=crcr]{%
-4.3702302362906	0.497550200803213\\
0.629769763709402	0.474532374100719\\
5.6297697637094	0.364343434343434\\
10.6297697637094	0.101062176165803\\
15.6297697637094	0.00703100775193798\\
20.6297697637094	0.000726456009913259\\
25.6297697637094	0.000140204271123491\\
30.6297697637094	5.95635715876197e-05\\
35.6297697637094	2.63596930263597e-05\\
40.6297697637094	1.5400616024641e-05\\
};

\addplot [smooth, color=mycolor3, mark=triangle*, mark options={solid, mycolor3}, forget plot]
  table[row sep=crcr]{%
-4.3702302362906	0.499287878787879\\
0.629769763709402	0.487833333333333\\
5.6297697637094	0.432666666666667\\
10.6297697637094	0.256327470686767\\
15.6297697637094	0.0576612021857923\\
20.6297697637094	0.0085238603988604\\
25.6297697637094	0.00230357973035797\\
30.6297697637094	0.00101683348498635\\
35.6297697637094	0.000581798375916023\\
40.6297697637094	0.00045577967416602\\
};

\addplot [smooth, color=mycolor4, mark=triangle*, mark options={solid, mycolor4}, forget plot]
  table[row sep=crcr]{%
-4.3702302362906	0.498035230352304\\
0.629769763709402	0.488979963570128\\
5.6297697637094	0.446810237203496\\
10.6297697637094	0.313117048346056\\
15.6297697637094	0.122795048934945\\
20.6297697637094	0.0326383981154299\\
25.6297697637094	0.013036858974359\\
30.6297697637094	0.00566557555919258\\
35.6297697637094	0.00350365415582807\\
40.6297697637094	0.00296127562642369\\
};

\addplot [smooth, dashed, color=mycolor1, mark=o, mark options={solid, mycolor1}, forget plot]
  table[row sep=crcr]{%
-4.37014015627852	0.493151515151515\\
0.629859843721482	0.464783333333333\\
5.62985984372148	0.259644808743169\\
10.6298598437215	0.0308308308308308\\
15.6298598437215	0.00284022038567493\\
20.6298598437215	0.000294206854716409\\
25.6298598437215	6.17902471609886e-05\\
30.6298598437215	1.8385732671447e-05\\
35.6298598437215	8.09552722121028e-06\\
40.6298598437215	3.34196674743086e-06\\
};

\addplot [smooth, dashed, color=mycolor2, mark=o, mark options={solid, mycolor2}, forget plot]
  table[row sep=crcr]{%
-4.37014015627852	0.49781124497992\\
0.629859843721482	0.470851318944844\\
5.62985984372148	0.342756132756133\\
10.6298598437215	0.122219343696028\\
15.6298598437215	0.0235891472868217\\
20.6298598437215	0.00377168525402726\\
25.6298598437215	0.000691736304549675\\
30.6298598437215	0.000200957470496549\\
35.6298598437215	9.34267600934268e-05\\
40.6298598437215	4.10016400656026e-05\\
};

\addplot [smooth, dashed, color=mycolor3, mark=o, mark options={solid, mycolor3}, forget plot]
  table[row sep=crcr]{%
-4.37014015627852	0.49930303030303\\
0.629859843721482	0.485637254901961\\
5.62985984372148	0.424320512820513\\
10.6298598437215	0.284032663316583\\
15.6298598437215	0.138060109289617\\
20.6298598437215	0.0418358262108262\\
25.6298598437215	0.0121966527196653\\
30.6298598437215	0.00343949044585987\\
35.6298598437215	0.0013938403644286\\
40.6298598437215	0.000877618308766486\\
};

\addplot [smooth, dashed, color=mycolor4, mark=o, mark options={solid, mycolor4}, forget plot]
  table[row sep=crcr]{%
-4.37014015627852	0.497479674796748\\
0.629859843721482	0.492194899817851\\
5.62985984372148	0.444494382022472\\
10.6298598437215	0.338664122137405\\
15.6298598437215	0.222081174438687\\
20.6298598437215	0.116342756183746\\
25.6298598437215	0.0542922008547009\\
30.6298598437215	0.02\\
35.6298598437215	0.00881704446921838\\
40.6298598437215	0.00622753733232093\\
};

\addplot[smooth,color=red,solid,mark=square*,mark options={solid, red}]
  table[row sep=crcr]{%
        1 2\\
};\label{P1}

\addplot[smooth, color=mycolor1, solid]
  table[row sep=crcr]{%
        1 2\\
};\label{P2}
\addplot[smooth, color=mycolor2, solid]
  table[row sep=crcr]{%
        1 2\\
};\label{P3}

\addplot[smooth, color=mycolor3, solid]
  table[row sep=crcr]{%
        1 2\\
};\label{P4}

\addplot[smooth, color=mycolor4, solid]
  table[row sep=crcr]{%
        1 2\\
};\label{P5}

\node [draw,fill=white, font=\tiny, anchor= south west, at={(-5,10^(-4.5))}]{
\shortstack[l]{
\ref{P1}~{Uncoded}\\
\ref{P5}~{Code Rate 3/4}\\
\ref{P4}~{Code Rate 2/3}\\
\ref{P3}~{Code Rate 1/2}\\
\ref{P2}~{Code Rate 1/3}}};

\end{axis}
\end{tikzpicture}%
\captionsetup{font=footnotesize}
	\caption{BER comparison between Gray and Advanced Mapping, $K_{cc}=3$}
	\label{fig:ber_mapping}
\end{figure}
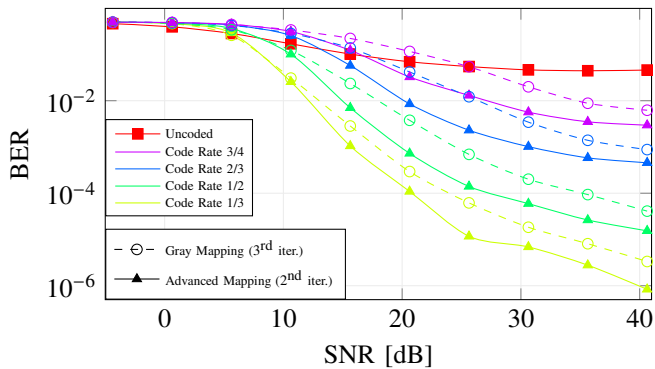
For the case with modulation order of $M_{\textrm{cpm}}=8$ and oversampling factor of $M=3$, channel coding is applied according to the extended discrete model shown in Fig.~\ref{fig:discrete_system}. The BER results are presented in Fig.~\ref{fig:ber_mapping}. The Gray and Advanced mapping is considered for the different code rates summarized in Table \ref{table:punctcodes1}, which corresponds to codes with constraint length $K_{cc}=3$. 
The results confirm that system designs using Advanced mapping outperform system designs using Gray mapping in terms of BER.
For systems with Gray mapping three iterations are considered and for systems with Advanced mapping only two, 
because no further performance gain was observed by increasing the iterations for the latter.
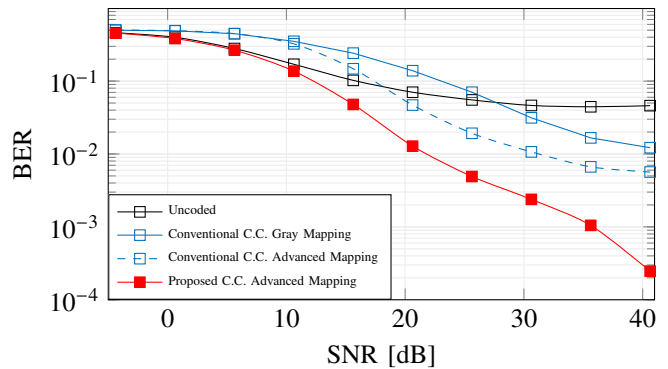
\begin{figure}[t]
	\centering
	\definecolor{mycolor1}{rgb}{0.00000,0.44706,0.74118}%
\begin{tikzpicture}

\begin{axis}[%
width=0.82\columnwidth,
height=0.4375\columnwidth,
scale only axis,
grid style={color=gray!15},
xmin=-5,
xmax=41,
xlabel={SNR [dB]},
ymode=log,
ymin=0.0001,
ymax=0.99,
yminorticks=true,
ylabel={BER},
xmajorgrids,
ymajorgrids,
yminorgrids,
legend style={at={(0,0)},anchor=south west,draw=black,fill=white,legend cell align=left,font=\tiny}
]
\addplot [smooth, color=black, mark=square, mark options={solid, black}]
  table[row sep=crcr]{%
-4.37244422025237	0.460915637860082\\
0.627555779747632	0.399605475040258\\
5.62755577974763	0.280030864197531\\
10.6275557797476	0.170204178537512\\
15.6275557797476	0.102389251997095\\
20.6275557797476	0.0703703703703704\\
25.6275557797476	0.0551764705882353\\
30.6275557797476	0.046343009676343\\
35.6275557797476	0.0444340844340844\\
40.6275557797476	0.046037037037037\\
};
\addlegendentry{Uncoded}



\addplot [color=mycolor1, mark=square, mark options={solid, mycolor1}]
  table[row sep=crcr]{%
-4.37095612757615	0.499259259259259\\
0.62904387242385	0.488913043478261\\
5.62904387242385	0.447277777777778\\
10.6290438724238	0.351343101343101\\
15.6290438724238	0.241816059757236\\
20.6290438724238	0.138802308802309\\
25.6290438724238	0.0703641456582633\\
30.6290438724238	0.0313599313599314\\
35.6290438724238	0.0166317016317016\\
40.6290438724238	0.0122033462033462\\
};
\addlegendentry{Conventional C.C. Gray Mapping} 


\addplot [smooth, dashed, color=mycolor1, mark=square, mark options={solid, mycolor1}]
  table[row sep=crcr]{%
-4.37158426797883	0.501150793650794\\
0.628415732021173	0.491293995859213\\
5.62841573202117	0.447412698412698\\
10.6284157320212	0.324371184371184\\
15.6284157320212	0.147586367880486\\
20.6284157320212	0.0469264069264069\\
25.6284157320212	0.0192296918767507\\
30.6284157320212	0.0106928356928357\\
35.6284157320212	0.00665834165834166\\
40.6284157320212	0.00565637065637066\\
};
\addlegendentry{Conventional C.C. Advanced Mapping} 


\addplot [smooth, color=red, mark=square*, mark options={solid, red}]
  table[row sep=crcr]{%
-4.37249978802042	0.455595238095238\\
0.627500211979577	0.383540372670807\\
5.62750021197958	0.265309523809524\\
10.6275002119796	0.137179487179487\\
15.6275002119796	0.047843137254902\\
20.6275002119796	0.0128210678210678\\
25.6275002119796	0.00491036414565826\\
30.6275002119796	0.00238524238524239\\
35.6275002119796	0.00104895104895105\\
40.6275002119796	0.000247104247104247\\
};
\addlegendentry{Proposed C.C. Advanced Mapping}

\end{axis}
\end{tikzpicture}%
\captionsetup{font=footnotesize}
	\caption{BER results for a code rate of $7/9$.} 
	\label{fig:ber_proposed}
\end{figure}
In Fig.~\ref{fig:ber_proposed}, the BER results for the proposed coding scheme are presented. The conventional coding scheme uses a convolutional code with rate $7/9$, generator polynomial $(5~7)$ and puncturing pattern  $1 1 \vert 0 1 \vert 0 1 \vert 1 0 \vert 1 0 \vert 0 1 \vert 1 1$. In contrast, the proposed coding scheme uses a convolutional code with rate $1/3$ for the third bit sub-channel as shown in Fig.~\ref{fig:prop_coding}(b).

\section{Conclusions}
\label{sec:conclusion}
Iterative detection and decoding applied to a CPM system with 1-bit quantization and oversampling at the receiver has been studied.
Different code rates have been considered and
it turns out that channel coding is beneficial in all the cases.
Additional performance gain can be achieved by using the proposed tailored bit mapping strategy in combination with a coding scheme that considers different binary sub-channels separately.



%




\ifCLASSOPTIONcaptionsoff
  \newpage
\fi



%
\bibliographystyle{IEEEtran}
\bibliography{bib-refs}

%








\end{document}